\documentclass[%
 reprint,
superscriptaddress,
showpacs,preprintnumbers,
 amsmath,amssymb,
 aps,
prb,
]{revtex4-1}
\usepackage{slashed}
\usepackage{graphicx}
\usepackage{dcolumn}
\usepackage{bm}

\begin{document}
\title{Angular dependence of the upper critical induction of  clean $s$- and $d_{x^2-y^2}$-wave superconductors with self-consistent  ellipsoidal effective
mass and Zeeman anisotropies}
\author{Aiying Zhao}\affiliation{Institute of Theoretical Physics, University of Science and Technology Beijing, Beijing 100083, China}
\affiliation{Department of Physics, University of Central Florida, Orlando, Florida, 32816-2385, USA}
\author{Qiang Gu}
\affiliation{Institute of Theoretical Physics, University of Science and Technology Beijing, Beijing 100083, China}
\email[Corresponding author: ]{qgu@ustb.edu.cn}
\author{Richard A. Klemm}
\affiliation{Department of Physics, University of Central Florida, Orlando, Florida, 32816-2385, USA}
\affiliation{U. S. Air Force Research Laboratory, Wright-Patterson Air Force Base, Ohio 45433-7251, USA}
\email[Corresponding author: ]{richard.klemm@ucf.edu.}

\date{\today}
\begin{abstract}
We employ the  Schr{\"o}dinger-Dirac method generalized to an ellipsoidal effective mass anisotropy
in order to treat the spin and orbital effective mass anisotropies self consistently, which is important
when Pauli-limiting effects on the upper critical field characteristic of singlet superconductivity
are present. By employing the Klemm-Clem transformations to map the equations of motion
into isotropic form, we then calculate the upper critical magnetic induction $B_{c2}(\theta, \phi, T)$ at
arbitrary directions and temperatures $T$ for isotropic $s$-wave and for anisotropic $d_{x^2-y^2}$ -wave
superconducting order parameters. As for  anisotropic $s$-wave superconductors, the reduced
upper critical field $b_{c2}$ is largest in the direction of the lowest effective mass, and is proportional to
the universal orientation factor  $\alpha(\theta,\phi)$. However, for  $d_{x^2-y^2}$-wave pairing, ${\bm B}_{c2}(\pi/2,\phi,T)$  exhibits either a four-fold pattern with $C_4$ symmetry just below the transition temperature $T_c$ that  rotates by $\pi/4$ as $T$ is lowered, or  a two-fold pattern with $C_2$ symmetry, depending upon the planar effective mass anisotropy.  This provides a new  method to distinguish these pairing symmetries in clean unconventional superconductors.
\end{abstract}

\maketitle

\section {Introduction}
Many unconventional superconductors such as the heavy-fermion superconductors, ferromagnetic superconductors, high $T_{c}$ copper oxide
superconductors, the iron-based superconductors, and magic-angle graphene superlattices have been found \cite{Sigrist1991,Cao2018,Cao2020}, and enormous efforts also have been triggered by the motivations to find higher
transition temperature $T_c$ and  higher upper critical field $H_{c2}$ superconductors,  and to understand the nature and  physical origin of the superconductivity in these
materials. Recently, the then-record high $T_c$ of 203 K was obtained in nominal H$_2$S under high pressure\cite{Drozdov}, exhibiting a conventional isotope effect. Since then, a variety of hydrides have been found under high pressure to be superconducting at least up to 243 K \cite{Pickard,Eremets}.  Since
critical field measurements on those rather conventional superconductors have not yet been made at low temperature $T$,  most of the presently more interesting and/or accessible materials have $T_c$ values below 100 K. These unconventional superconductors usually have strongly anisotropic structures and  upper critical fields that exhibit obvious and
unique orientational dependencies. Measurement of the upper critical induction $B_{c2}=\mu_0H_{c2}$ in non-magnetic systems is an important tool to
understand various superconductors, since it is a thermodynamic measurement from which coherence lengths, anisotropy parameters, and pair-breaking mechanisms can be extracted.

In unconventional superconductors, ${\bm B}_{c2}$ usually exhibits obvious and unique orientational dependencies. One of the reasons is that the Fermi surface (FS) is not always a single closed surface, and sometimes more than one electronic orbital can contribute to the overall FS, so that multi-band effects should be taken into account\cite{Gurevich2003,Gurevich2010,Kogan2012}. Even FS reconstruction can occur in hole-doped cuprate superconductors\cite{Norman2010}, which could be due to the onset of a charge- or spin-density wave, as are known to occur respectively in the transition metal dichalcogenides and certain organic layered superconductors\cite{Klemm2000,Klemm2015,Layered}. But it is the simplest and most convenient approach at present to theoretically assume that the FS is a sphere for isotropic superconductors, ellipsoidal for anisotropic superconductors or rippled cylindrical for layered superconductors\cite{Klemm1975,Layered,Wang2016}. We approximate the FS structure as an ellipsoid with three different effective masses and do not consider either multi-band effects or FS reconstruction in this work \cite{Aiying2021}.

In addition to FS anisotropy, the Pauli paramagnetic (or Zeeman) energy can also affect ${\bm B}_{c2}$. Although for triplet pair-spin superconductors
containing all three spin states, there is no paramagnetic suppression of the superconductivity if the external field is orthogonal to the direction along
which the Cooper pairs have zero total spin, a magnetic field parallel to the direction of zero total spin will be pair breaking\cite{Mineev,Aoki,ScharnbergKlemm1980,ScharnbergKlemm1985,Loerscher,ZhangSr2RuO4,Zhangfirstorder}. But this is not the case for a singlet pair
state. It always maintains the singlet state, no matter which direction the state is observed, and ${\bm B}_{c2}$ can be bounded by the paramagnetic
effect for all field directions, provided that its temperature $T$ slope at $T_c$ is sufficiently negative and that counteracting effects such as spin
fluctuations and spin-orbit scattering are sufficiently weak. Therefore, it is necessary to add the paramagnetic term to study singlet superconductors.
However, many workers that developed theories to fit the experimental data have neglected this term or have added it by hand, without making it consistent
with the intrinsic effective mass anisotropy appropriate for the material under study. In this work, the most important point is that we study the effects of the self-consistently anisotropic Zeeman energy and the orbital effective mass anisotropy  from the anisotropic non-relativistic limit, that we derived previously \cite{Aiying2021}

Generally, there exist two distinct ways to induce pair-breaking of unconventional superconductors by an applied magnetic field, i.e., orbital \cite{Helfand1966} and
spin-paramagnetic effects \cite{Clogston1962,Chandrasekhar1962}. The actual $H_{c2}$ of real materials depends upon both of these effects, both of which can be greatly modified by the same effective mass anisotropy parameters. For an isotropic superconductor, the relative importance of the orbital and spin-paramagnetic effects is described by the Maki parameter\cite{Maki1966}, $\alpha_{M}=\sqrt{2} H_{c2}^{orb}(0)/H_{c2}^{P}(0)$. Since $\alpha_M$ is known to be on the order of $\Delta(0)/E_{F}$, where $E_{F}$ is the Fermi energy, $\alpha_{M}\ll 1$, indicating that the paramagnetic effect is negligibly small in most
superconductors. However, in layered superconductors, or in heavy fermion superconductors that have a quite small $E_F$, $\alpha$ can be larger than
unity. Or in ''Ising'' layered superconductors, for which $B_{c2}$ parallel to the layers can well exceed the Pauli paramagnetic limit for an isotropic superconductor, the paired spins are less susceptible to Pauli pair-breaking due to the effective internal Zeeman-like field\cite{Lu2015}. But in nearly all of the calculations to date on  superconductors with anisotropic FSs, the Zeeman energy was assumed to completely independent of the FS anisotropy parameters.  As shown in the following, such assumptions violate the laws of special relativity, and must be completely revised for anisotropic materials.

We first note that beyond those two primary pair-breaking mechanisms,  the upper critical field is
affected by many additional factors: magnetic structure that may coexist or interfere with superconductor\cite{Aoki}, spin-orbit scattering\cite{Klemm1975,Werthamer1966}, flux pinning, the possibility of a Fulde-Ferrell-Larkin-Ovchinnikov (FFLO) phase\cite{Fulde1964,Larkin1964,Gruenberg1966,Matsuda2007,Agosta2017}, structural asymmetry\cite{Yip2014,Kneidinger2015,Goh2012}, and the spin and orbital symmetries of the superconducting order parameter. For example, in $p$-wave superconductors, not only does the triplet pair-spin configuration allow for strong violations of the Pauli limit\cite{Mineev,Aoki,ScharnbergKlemm1980,ScharnbergKlemm1985,Loerscher,ZhangSr2RuO4,Zhangfirstorder}, as in the case of URhGe, the reentrant phase above $B_{c2}$ exceeds the Pauli limit by a factor of 20, there are profound differences in the temperature dependence of $B_{c2}$ for the field along the single-axis pairing direction from those perpendicular to it\cite{Aoki,ScharnbergKlemm1980,ScharnbergKlemm1985,Loerscher}. Much more subtle order parameter anisotropy effects can
still arise for singlet spin superconductors, especially if strong spin-orbit scattering\cite{Werthamer1966,Klemm1975} or Ising single-spin direction
locking\cite{Lu2015} are present.

In addition, as is well known in the high-$T_c$ community, there has been an ongoing battle regarding the orbital symmetry of the superconducting order parameter in the
cuprates\cite{Shen2003,Tsuei2000,MoessleKleiner,Li1999,Takano2002,Takano2003,Latyshev,Klemm2005,Tsinghua2,Korea,Harvard}. The difficulties in such determinations are complicated
by the ubiquitous presence of a ``pseudogap'', which is most likely either a charge-density wave (CDW), as occurs, even with a node, in the hole bands of $2H$-TaS$_2$, one of the transition
metal dichalcogenides\cite{Layered,Klemm2015,Tonjes}, or a spin-density wave, as can occur in the organic layered superconductors and the pnictides, or possibly a combination of
both\cite{Klemm2000}. In the case of $2H$-TaS$_2$, the nodal CDW disappears upon intercalation, so it requires more than one layer to exist\cite{Gamble}.  It was recently found that in a single crystal of Bi2212 cleaved in vacuum and covered with a monolayer of CuO$_2$ to protect the surface against reacting with the remaining atoms, {\it etc.} in the vacuum, before scanning tunnelling microscopy (STM) studies were performed\cite{Tsinghua}.  Those authors found that the Bi2212 pseudogap and the superconducting order
parameter separate into rather distinct spatial nanodomains.  The pseudogap was found by STM studies to have a node that persists up to its onset well above the superconducting $T_c$, but the STM-observed superconducting gap was
nodeless and essentially isotropic, with a flat STM bottom and large coherence peaks, that disappears at or near to $T_c$\cite{Tsinghua}. We note that the Postech and Harvard groups were unaware of the nodal ``pseudogap'' (most likely a CDW) in underdoped and optimally doped Bi2212\cite{Klemmprivate}, but which disappears in the overdoped region of the Bi2212 phase diagram, as was studied in the original $c$-axis twist experiment on that compound\cite{Li1999}.  This nodal CDW has confused all previous angle resolved photoemission experiments that average over an area of about 1 mm$^2$, which covers thousands of such nanodomains\cite{Shen2003}.

 In addition, the narrow linewidth of the emission averaged over all directions at 2.4 THz from a disk-shaped device of Bi2212 is difficult to explain unless the quasiparticle energy has a minimum gap of at least 9.8 meV \cite{Kashiwagi,Kleinerprivate}.  Hence, these new experiments bring the published interpretations of many previous experiments into serious question, and new and repeated tests of
order parameter symmetry on better quality  samples are strongly warranted.  Here we show that for clean anisotropic type-II superconductors, the angular and temperature dependence of the upper critical field $H_{c2}$ can provide additional evidence for the orbital symmetry of the superconducting order parameter.

Recently, the number of clean layered and two-dimensional (2D) superconductors has increased substantially\cite{Cao2018,Cao2020}.  In addition, a model based upon coherent Josephson $c$-axis tunnelling was studied for the $d_{x^2-y^2}$-wave order parameter neglecting non-local effects\cite{Wang2016}, finding some interesting azimuthal $\phi$ angular dependencies of $B_{c2}(\theta,\phi,T)$. But that model was only marginally applicable to the  cuprate  YBa$_2$Cu$_3$O$_{7-\delta}$, for which $B_{c2}(\theta,\phi,T)$ studies have been made or attempted\cite{Naughton,Welp1,Welp2,Miura}, as detwinned samples have an in-plane effective mass anisotropy of about a factor of 2, depending upon the precise oxygen stoichiometry. More recent interesting $B_{c2}(\theta,\phi,T)$ measurements on CeCu$_2$Si$_2$ were also made\cite{CeCu2Si2}.  Since those pioneering measurements of $B_{c2}(\theta,\phi,T)$, some of the world's high magnetic field laboratories have improved their capabilities, and can now operate at sufficiently low $T$ ranges with dc applied fields as large as 45 T, so that accurate experimental measurements of  $B_{c2}(\theta,\phi,T)$ can now be obtained well below $T_c$ in materials of interest.  Since a complete consensus on $c$-axis Josephson tunnelling experiments on hole-doped cuprates has not yet been attained, and the in-plane Josephson tunnelling experiments \cite{Tsuei2000} have not yet been successfully explained in a manner consistent with the majority of the $c$-axis Josephson experiments, other experiments to test the orbital symmetry of the superconducting order parameter in unconventional superconductors are warranted.  But since it was recently shown that a proper relativistically consistent treatment of the orbital and spin effective mass anisotropies can be made\cite{Aiying2021}, such experiments should also be relativistically consistent with those properties.

This paper is organized as follows. In Section \uppercase\expandafter{\romannumeral2}, the non-relativistic limit of the Hamiltonian in an orthorhombically anisotropic medium is presented. The superconducting pairing interactions are also presented, and the general gap functions for all reduced temperatures $t=T/T_c$  for $s$-wave and $d_{x^2-y^2}$-wave are derived. The numerical results based on the equations presented in Section \uppercase\expandafter{\romannumeral2} are presented and discussed in Section \uppercase\expandafter{\romannumeral3}. Details of the effects of the self-consistent orbital effective mass and Zeeman energy anisotropies combined with the two order parameter anisotropies are presented and discussed. Finally, the main results are presented in \uppercase\expandafter{\romannumeral4}.

\section{The Model}
\subsection{The Transformation}
\begin{figure}
\center{\includegraphics[width=0.45\textwidth]{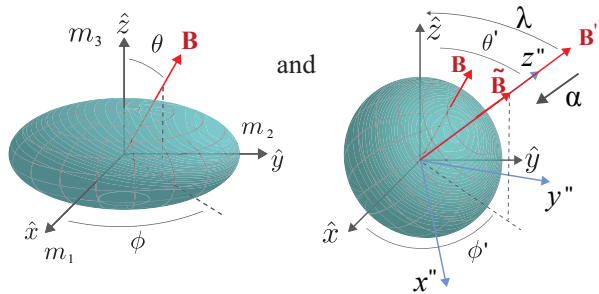}
\caption{(Color online) Shown is a schematic diagram of the transformation.}
\label{fig1}}
\end{figure}

We begin  the non-relativistic limit of in an orthorhombically anisotropic medium
\begin{eqnarray}
H_0 &=&\Bigl[\sum_{\mu=1}^3 \frac{1}{\sqrt{2m_{\mu}}} \sigma_{\mu}(-i\partial_{\mu} + e A_{\mu})\Bigr]^2,
\end{eqnarray}

where ${\bm B}(\bm{r})={\bm\nabla}\times{\bm A}(\bm{r})$ is the magnetic induction, ${\bm{A}}$ is the vector potential, $e$ is the absolute value of the
electron charge, the $\sigma_{\mu}$ are the Pauli matrices, ${\partial_{\mu}\equiv\partial/ \partial x_{\mu}}$, $x_{\mu}$ is $x$, $y$, $z$ for $\mu=1, 2,
3$, respectively, and we set $\hbar=k_B=c=1$. At $B_{c2}$, the supercooling field, or the border  between  the superconducting and normal state of a
non-magnetic metal, we assume that the transition is a second-order thermodynamic transition, and that first-order transitions to some other state, such as the FFLO state\cite{Fulde1964,Larkin1964,Gruenberg1966,Matsuda2007,Agosta2017}, do not occur in the ${\bm B}(T)$ range of the experiments. In addition, we assume that a competing CDW or SDW state is either absent or does not contribute to the ${\bm B}_{c2}=\mu_0{\bm H}_{c2}$ measurements. Although such competing order parameters have apparently distorted the results of phase-sensitive $c$-axis twist experiments in underdoped or optimally-doped Bi2212 \cite{Korea,Harvard}, $B_{c2}(T)$ curves in that material can only be measured close to $T_c$ \cite{Naughton}, and the transitions are so broad at lower $T$ values as to make meaningful estimates of ${\bm B}_{c2}(T)$ almost impossible\cite{Layered}. We therefore may assume  $\bm{B}$ to be constant throughout the superconductor, writing
\begin{eqnarray}
{\bm{B}}& =&B(\sin\theta\cos\phi, \sin\theta\sin\phi,\cos\theta)\nonumber
\end{eqnarray}
in spherical coordinates.
Expanding the squared quantities in $H_0$, we obtain \cite{Aiying2021}
\begin{eqnarray}
H_0&=&\sum_{\mu}\Bigl[\frac{(-i\partial_{\mu} + eA_{\mu})^2}{2m_{\mu}}+\frac{e}{2 m_g} \sigma_{\mu}\sqrt{\overline{m}_{\mu}} B_{\mu}\Bigr],
\end{eqnarray}
where $m_g=(m_1m_2m_3)^{1/3}$ is the geometric mean effective mass, $\overline{m}_{\mu}=m_{\mu}/m_g$, and $\sigma_{\mu}$ is the matrix representing the $\mu^{\rm th}$ component of the electron spin and $\tilde{\bm r}=\alpha{\bm r}''$.  We emphasize that the effective masses $m_{\mu}$ appear in the orbital part of the electron motion, and are not necessarily related to the rest mass $m$ of an electron in vacuum.

We then make the Klemm-Clem transformations\cite{Layered,Klemm1980}, and Fig.1 gives the schematic diagram of the transformation, after which
 $H_{0}$ reduces to
\begin{eqnarray}
\tilde{H}_{0}&=&\frac{1}{2m_g}(-i{\bm\nabla}^{\prime\prime} + e{\bm A}^{\prime\prime})^2 + \alpha \frac{e}{2 m_g}  \sigma_z\tilde{B}_z,\\
\alpha(\theta,\phi)\!\!&=&\!\!\![\overline{m}_1\sin^2\theta\cos^2\phi+\overline{m}_2\sin^2\theta\sin^2\phi+\overline{m}_3\cos^2\theta]^{1/2},\label{alpha}\nonumber\\
\end{eqnarray}
where $\tilde{B}_z=B_z''/\alpha(\theta,\phi)$.
We note that  the transformed Zeeman energy is proportional to $\alpha(\theta,\phi)$,  the direction of the magnetic field ${\bm B}({\theta,\phi})$ has changed to $({\theta^{\prime},\phi^{\prime}})$,  the Laudau-level orbits scale equivalently with the transformed Fermi Surface, and the geometric mean effective mass is $m_g$.

We have quantized the rotated spins along  $z^{\prime\prime}$ axis, the direction of the transformed $\tilde{\bm B}$.  Thus, in second quantization, the full
Hamiltonian $H'=H-\mu N$ may now be written in transformed real space  as
\begin{eqnarray}
H'&=&\sum_{\sigma=\pm}\int d^3{\bm r}^{\prime\prime}  \biggl[\psi^{\dag}_{\sigma}({\bm r}^{\prime\prime})
\Bigl[\frac{1}{2m_g}\Bigl(-i{\bm\nabla}^{\prime\prime}+e{\bm A}^{\prime\prime}(\tilde{\bm r})\Bigr)^2\nonumber\\
& &\hskip15pt-\mu+ \alpha \frac{e}{2 m_g}\sigma_z  \tilde{B}_z\Bigr]\psi_{\sigma}({\bm r}^{\prime\prime})\nonumber\\
&&\!+\frac{1}{2}\!\int\!d^3{\bm r_1}^{\prime\prime} \psi^{\dag}_{\sigma}({\bm r_2}^{\prime\prime}) \psi^{\dag}_{-\sigma}({\bm r}_{1}^{\prime\prime})
V({\bm r}_{2}^{\prime\prime}-{\bm r}_{1}^{\prime\prime}) \psi_{-\sigma}({\bm r}_{1}^{\prime\prime})\psi_{\sigma}({\bm r}_{2}^{\prime\prime})\biggr],\nonumber\\
\end{eqnarray}
where $\mu$ is the chemical potential, $\tilde{\bm r}=\alpha{\bm r}''$, and the untransformed pairing interactions in reciprocal space may be written as\cite{ScharnbergKlemm1980}
\begin{eqnarray}
V(\hat{\bm k},\hat{\bm k}')&=&\left\{\begin{array}{lc} V_0,& s\\
V_0(\hat{k}_x^2-\hat{k}_y^2)(\hat{k}_x^{'2}-\hat{k}_y^{'2}),&d_{x^2-y^2}\end{array}\right.\label{interaction}
\end{eqnarray}
We note that the anisotropic scale transformation also causes the magnitude and direction of $\bm{k}(\theta_k,\phi_k)$ to be changed into $k^{\prime}(\theta^{\prime},\phi^{\prime})$, since $k_{\mu}'=\frac{k_{\mu}}{\sqrt{\bar{m}_{\mu}}}$, and the overall magnitude of the wave vector changes according to $k'=\beta(\theta_{k},\phi_{k}) k$, where
$\beta(\theta_k,\phi_k)$ is given by
\begin{eqnarray}
\beta=&\Bigl[\frac{\sin^2 \theta_{k} \cos^{2}\phi_{k}}{\overline{m}_{1}}+\frac{\sin^2\theta_{k}\cos^2\phi_{k}}{\overline{m}_{2}}
+\frac{\cos^2\theta_{k}}{\overline{m}_{3}}\Bigr]^{1/2},\label{beta}
\end{eqnarray}
and the angle relationship is
\begin{eqnarray}
&&\sin\theta_{k^{\prime}}=\frac{\beta(\frac{\pi}{2},\phi_k)}{\beta(\theta_{k},\phi_k)} \sin\theta_k;
\cos\theta_{k^{\prime}}=\frac{1/\sqrt{\overline{m}_3}}{\beta(\theta_{k},\phi_k)} \cos\theta_k; \nonumber\\
&&\sin\phi_{k^{\prime}}=\frac{1/\sqrt{\overline{m}_2}}{\beta(\frac{\pi}{2},\phi_k)} \sin\phi_k;
\cos\phi_{k^{\prime}}=\frac{\beta(1/\sqrt{\overline{m}_1}}{\beta(\frac{\pi}{2},\phi_k)} \cos\theta_k; \nonumber\\
\end{eqnarray}
However, at the chemical potential (or  $E_F$), we have $(k')^2=2m_g\mu$, so in integrations about the Fermi energy, we may safely approximate
\begin{eqnarray}
\beta^2k^2\Big|_{\mu}&\approx&2m_g\mu.
\end{eqnarray}

\subsection{Gap Function}

All the representation of physical quantities are shown in new rectangular coordinate system with double prime, so it is convenient to omit the double prime in the following. We define the superconducting order parameter to be
\begin{eqnarray}
\Delta_{\sigma,\sigma'}(\bm{r}, \bm{r}')&=& V(\bm{r}, \bm{r}')F_{\sigma,\sigma'}(\bm{r}, \bm{r}',0^{+}),
\end{eqnarray}
in which $V(\bm{r}, \bm{r}')$ is the pairing interaction in weak-coupling theory. It may be written as $V(\bm{r}, \bm{r}')= \lambda \delta(\bm{r}-\bm{r}')$, and then $\Delta_{\sigma,\sigma'}(\bm{r},\bm{r}')=\Delta(\bm{r})\delta(\bm{r}-\bm{r}')\delta_{\sigma,-\sigma'}$, for traditional $s$-wave pairing.  For
non-$s$-wave superconductors, it is easier to define the gap function in wave vector space\cite{ScharnbergKlemm1980}.
Using Gor'kov's description of weakly coupled superconductors, we can derive the fully transformed equations of motion for the normal and anomalous
temperature Green functions of anisotropic superconductors\cite{AGD},
\begin{eqnarray}
&&\Bigl[i\omega_n-\frac{({\bm\nabla}/i - e{\bm A})^2}{2m_g}+E_{F} - \frac{e}{2 m_g}\alpha\sigma_z\tilde{B}_z\Bigr]G_{\sigma,
\sigma'}({\bm{r}},{\bm{r}'},\omega_{n}) \nonumber \\
&&+\sum_{\rho}\int d^{3} {\bm \xi}\Delta_{\sigma, \rho}({\bm{r}},{\bm \xi})F_{\rho,
\sigma'}^{\dagger}({\bm\xi},{\bm{r}'},\omega_n)=\delta_{\sigma,\sigma'}\delta({\bm{r}}-{\bm{r}'}),\nonumber\\
\end{eqnarray}
\begin{eqnarray}
&&\Bigl[-i\omega_n-\frac{(i{\bm\nabla} - e{\bm A})^2}{2m_g}+E_{F} - \alpha\frac{e}{2 mg} \sigma_z\tilde{B}_z\Bigr]F_{\sigma,
\sigma'}^{\dagger}({\bm{r}},{\bm{r}'},\omega_{n}) \nonumber \\
&&-\sum_{\rho}\int d^{3} {\bm\xi} \Delta_{\sigma, \rho}^{*}({\bm{r}}, {\bm{\xi}})G_{\rho, \sigma'}({\bm{\xi}},{\bm{r}'}, \omega_n)=0,
\end{eqnarray}
where $\omega_n=(2n+1)\pi T$ are the Matsubara frequencies.
For a charged anisotropic superfluid in a magnetic field. We use the method introduced by Scharnberg  and Klemm in their study of  $p$-wave
superconductors to obtain the gap function before the transformations, \cite{ScharnbergKlemm1980}
\begin{eqnarray}
\Delta_{\sigma, \sigma'}(\bm{R},\bm{k}_{F})\!&=&\!2\pi TN(0)\sum_{\omega_{n}}\int\frac{d\Omega_{\bm{k}'}}{4\pi} V(\bm{k}_{F} -
\bm{k}'_{F})\nonumber\\
&&\times\int_{0}^{\infty}d\xi e^{-2|\omega_{n}|\xi}\cos[\frac{e}{2 m_g}\alpha(\sigma_z - \sigma'_{z})B_{z}\xi] \nonumber \\
&&\times e^{-i{\rm sgn}(\omega_{n})\xi \bm{v}_{F}({\bm{\hat{k}}}_{F}^{'})\cdot{\bm\Pi}({\bm
R})}\Delta_{\sigma,\sigma^{'}}(\bm{R},\bm{k}_{F}'),\nonumber\\
\end{eqnarray}
where ${\bm\Pi}({\bm R})={\bm\nabla}_{\bm{R}}/i+2e{\bm A}({\bm R})$, $\bm{R}$ denotes the  center of mass position of the paired electrons, and $N(0)$ is
the electronic density of states.
\subsection{Particular Models}

The most notable quality of an anisotropic superconductor, especially for a heavy fermion superconductor (HFS), is its anisotropic gap. Such as anisotropic
$p$-wave superconductors, one may have a nodeless gap, or a gap with either planar nodes (such as a polar state), or point nodes (such as an axial state),
where the gap vanishes on the Fermi surface. We consider $s$-wave and $d_{x^2-y^2}$-wave superconductors with ellipsoidally anisotropic masses and
Zeeman energies in this section.

For a conventional $s$-wave superconductor, the gap function $\Delta_s(\bm{R},\bm{\hat{k}}_{F})=\Delta_s(\bm{R})$ and the interaction
$V(\hat{\bm{k}},\hat{\bm{k}}^{'})=V_{0}$ is a constant.  But for a $d_{x^2-y^2}$ superconductor, the interaction in Eq. (\ref{interaction}) leads to
\begin{eqnarray}
\Delta_d(\bm{R},\bm{\hat{k}}_{F})&=&\Delta_d(\bm{R})(\hat{\bm{k}}_{x}^{2}-\hat{\bm{k}}_{y}^{2}),
\end{eqnarray}
in which $\hat{\bm{k}}_F$ denotes the unit vector of the wave vector ${\bm k}_F$ on the Fermi surface, and the $\hat{k}_i$ are its components.

We then transform the unit wave vector components, $\hat{\bm{k}}_i(\theta_{k}, \phi_{k})$  into the $\hat{{\bm{k}}}^{\prime\prime}_i(\theta_{k^{\prime}},\phi_{k^{\prime}})=\alpha\hat{\tilde{k}}_i(\theta_{k^{\prime}},\phi_{k^{\prime}})$, obtaining
\begin{eqnarray}
\hat{k}_{x}&=&
\sqrt{\overline{m}_1}\alpha\beta(\cos\theta'\cos\phi'\hat{\tilde{k}}_x
-\sin\phi'\hat{\tilde{k}}_y+\sin\theta'\cos\phi'\hat{\tilde{k}}_z),\nonumber\\
\hat{k}_{y}&=&\sqrt{\overline{m}_2}\alpha\beta(\cos\theta'\sin\phi'\hat{\tilde{k}}_x
+\cos\phi'\hat{\tilde{k}}_y+\sin\theta'\sin\phi'\hat{\tilde{k}}_z), \nonumber \\
\hat{k}_{z}&=&\sqrt{\overline{m}_3}\alpha\beta( - \sin\theta'\hat{\tilde{k}}_x
+\cos\theta'\hat{\tilde{k}}_z),
\end{eqnarray}
and the $d_{x^2-y^2}$ pairing interaction transforms according to
\begin{eqnarray}
V(\hat{\bm k},\hat{\bm k}^{'})&=&V_0f({\bm k})f({\bm k}')\\
f({\bm k})&=&(\hat{k}_{x}^{2}-\hat{k}_{y}^{2})\rightarrow\alpha^2\beta^2f(\tilde{\bm k}),
\end{eqnarray}
where
\begin{eqnarray}
f(\tilde{\bm k})\!\!&=&\!\overline{m}_{1}\Bigl(\cos\theta'\cos\phi'\hat{\tilde{k}}_{x} - \sin\phi'\hat{\tilde{k}}_{y}
+\sin\theta'\cos\phi'\hat{\tilde{k}}_{z}\Bigr)^{2}\nonumber\\
&&\!\!-\overline{m}_2\Bigl(\cos\theta'\sin\phi'\hat{\tilde{k}}_x+\cos\phi'\hat{\tilde{k}}_y+\sin\theta'\sin\phi'\hat{\tilde{k}}_z\Bigr)^2\>\>\nonumber\\
\end{eqnarray}
It is then elementary to transform the gap function
\begin{eqnarray}
\Delta_d(\bm{R},\bm{\hat{k}}_{F})&=&\Delta_d(\bm{R})f({\bm k})\\
&\rightarrow&\Delta_d(\tilde{\bm{R}})\alpha^2\beta^{2}f(\tilde{\bm k}),
\end{eqnarray}
where $f(\tilde{\bm k})$ is given by Eq. (18).
\begin{figure*}
\begin{center}
\includegraphics[width=0.3\linewidth]{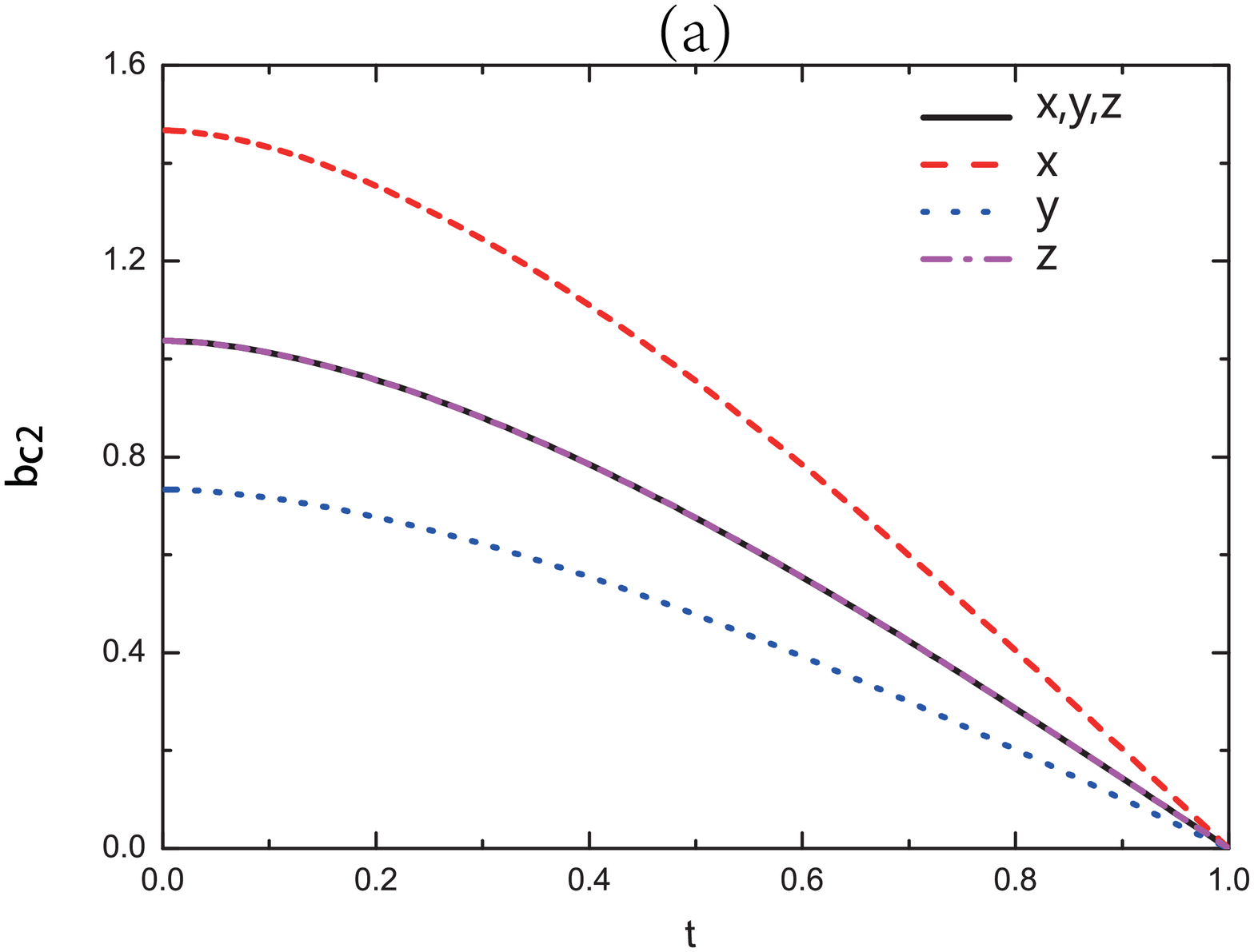}
\includegraphics[width=0.3\linewidth]{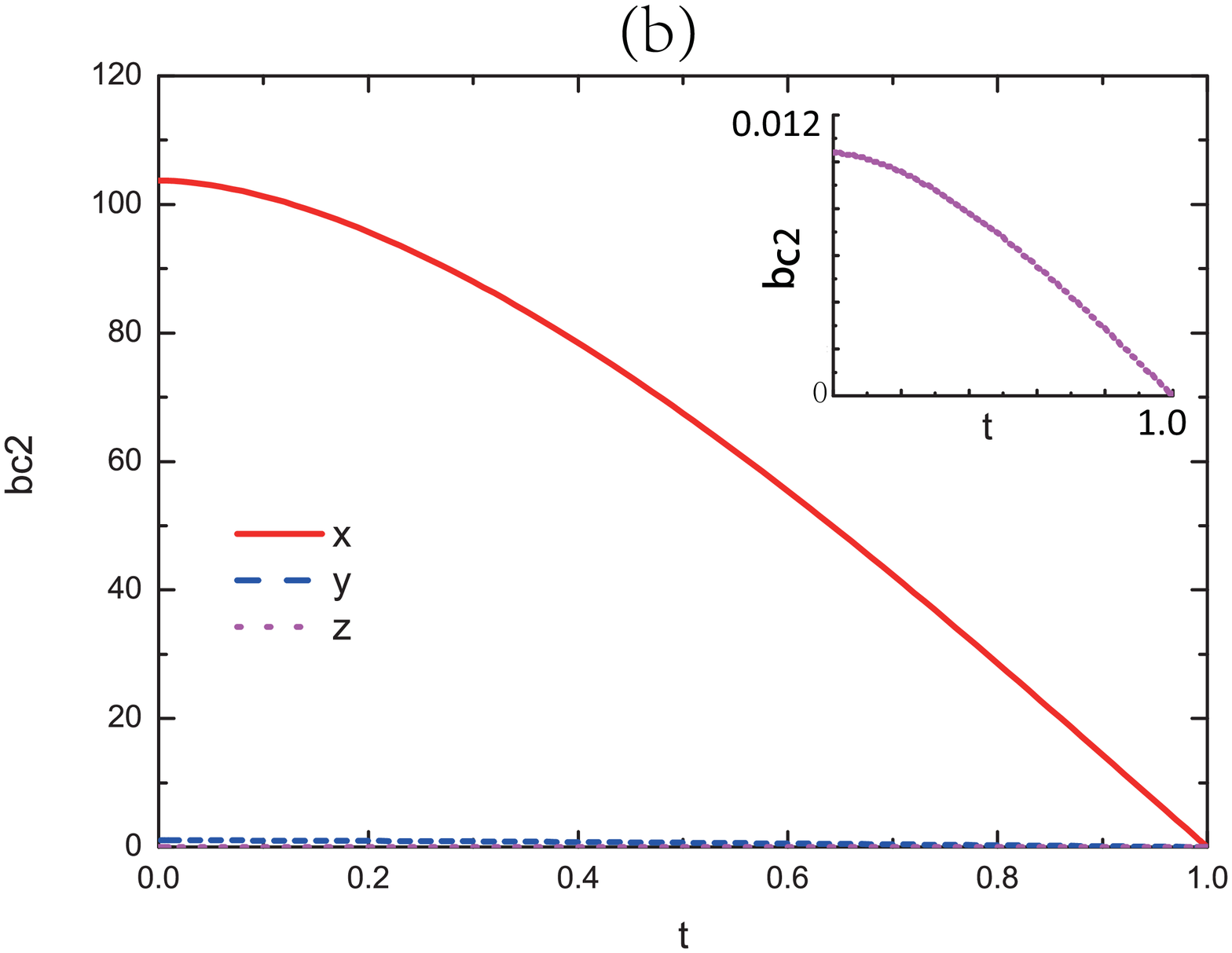}
\includegraphics[width=0.3\linewidth]{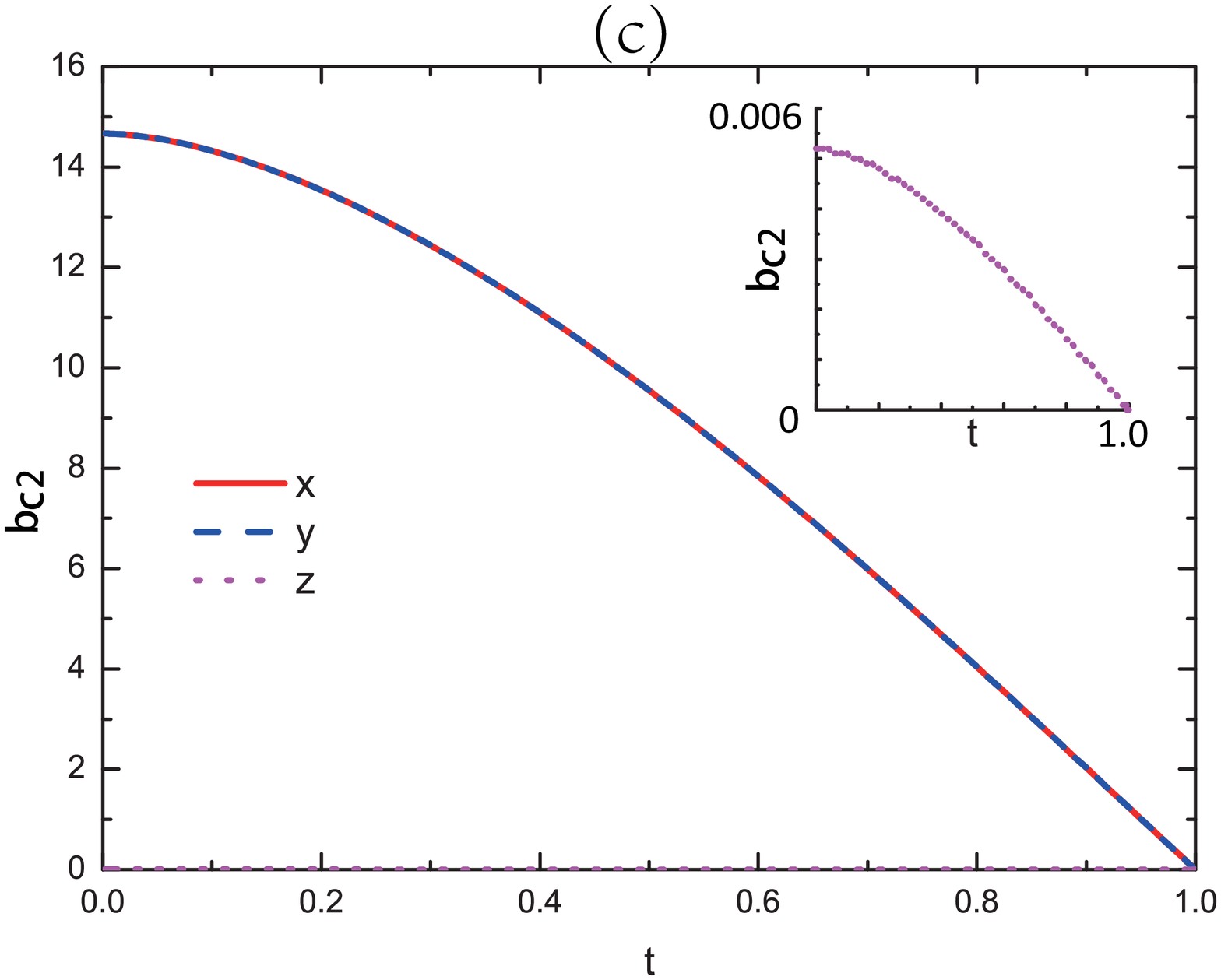}
\vspace{-5mm}
\end{center}
\caption{(Color online) Plots of the reduced upper critical field $b_{c2}(t)$ defined in Eq. (\ref{bc2}) for an $s$-wave superconductor with an anisotropic FS characterized by different effective masses and a correspondingly anisotropic Zeeman energy. The isotropic $(\overline{m}_1=\overline{m}_2=\overline{m}_3=1)$ case is identical to the solid black curve. For an $s$-wave OP with an anisotropic effective mass characterized by $\overline{m}_1=0.5$, $\overline{m}_2=2.0$, and $\overline{m}_3=1.0$, $b_{c2}(t)$ curves for the field along the $\hat{x}$ (dashed red), $\hat{y}$ (dotted blue), and $\hat{\bm z}$ (dash dot magenta) directions, respectively, are also shown in (a). Plots of the reduced $b_{c2}$ for a highly anisotropic effective mass with strong spin-orbit interactions for an $s$-wave OP are shown in (b)and (c). In both figures, ${\bm b}_{c2}$ along the $x$ direction is in solid red, along the $y$ direction is in dashed blue (just above the horizontal axis), and along the $z$ direction is in dotted magenta shown in the insets. In (b), $\overline{m}_1=1.0\times10^{-4}$, $\overline{m}_2=1.0$, and $\overline{m}_3=1.0\times10^4$, in (c), $\overline{m}_1=\overline{m}_2=0.0050$, and $\overline{m}_3=4.0\times10^4$ }
\label{fig2}
\end{figure*}

For respective  $s$-wave and $d_{x^{2}-y^{2}}$-wave order parameters, we have
\begin{eqnarray}
\Delta_s({\bm{R}})&=&2\pi TN(0)\sum_{\omega_{n}}\int\frac{d\Omega_{\bm{k}'}}{4\pi} V_{0} \int_{0}^{\infty}d\xi
e^{-2|\omega_{n}|\xi} \nonumber\\
& &\!\!\times \cos(2\frac{e}{2 m_g}\alpha\tilde{B}\xi)e^{-i{\rm sgn}(\omega_{n})\xi{\bm v}_{F}({\bm{\hat{k}}}_{F}^{'})\cdot{\bm\Pi}({\bm R})}
\Delta_s(\tilde{\bm{R}}),\nonumber\\
\Delta_d({\bm{R}})&=&2\pi TN(0)\sum_{\omega_{n}}\int\frac{d\Omega_{\bm{k}'}}{4\pi} V_{0}\alpha^2\beta^2f^2(\tilde{\bm
k}')\int_{0}^{\infty} d\xi \nonumber\\
 &&\!\!\times
 e^{-2|\omega_{n}|\xi}\cos(2\frac{e}{2 m_g}\alpha\tilde{B}\xi)e^{-i{\rm sgn}(\omega_{n})\xi \bm{v}_{F}({\bm{\hat{k}}}_{F}^{'})\cdot{\bm\Pi}({\bm
 R})}\nonumber\\
 & &\times \Delta_d({\bm{R}}),
\end{eqnarray}
where
\begin{eqnarray}
{\bm\Pi}({\bm R})&=&-i\alpha{\bm\nabla}_{{\bm R}}+2e{\bm A}({\bm R}).
\end{eqnarray}
As shown previously\cite{ScharnbergKlemm1980,Loerscher},  the gap functions $\Delta_i({{\bm R}})$ may be expanded
\begin{eqnarray}
\Delta_i({\bm{R}})&=&\sum_{n}a_{n,i}|n({\bm R})>,
\end{eqnarray}
for $i=s,d$
in terms of the complete orthonormal set of one-dimensional harmonic oscillator eigenfunctions $|n\rangle$, where
 $<n'|n>=\delta_{n,n'}$,
we respectively obtain a closed expression for the $a_{n,s}$  and a recursion relation for the $a_{n,d}$,
\begin{eqnarray}
a_{n,s}&=&a_{n,s}2\pi TN(0)\sum_{\omega_{m}}\int\frac{d\Omega_{\bm{k}^{'}}}{4\pi} V_{0}\int_{0}^{\infty} d\xi
e^{-2|\omega_{m}|\xi}\nonumber\\
& &\times \cos(2\frac{2}{2 m_g}\alpha\tilde{B}\xi)e^{-\frac{1}{2}|\eta|^2}\sum_{n'=0}^{\infty}L_{n'}^{0}(|\eta|^2),\\
a_{n,d}&=&2\pi TN(0)\sum_{\omega_{m}}\int\frac{d\Omega_{\bm{k}^{'}}}{4\pi} V_{0}\beta^2f^2({\bm k}') \nonumber\\
&&\times\sum_{n^{'}=0}^{\infty}\int_{0}^{\infty}\frac{d\xi}{\alpha} e^{-2|\omega_{m}|\xi}
\cos(2\frac{e}{2 m_g}\alpha\tilde{B}\xi)\nonumber\\
& &\times a_{n',d}<n({\bm R})| e^{-i{\rm sgn}(\omega_{m})\xi \bm{v}_{F}({\bm{\hat{k}}}_{F}^{'})\cdot{\bm\Pi}({\bm R})}|n'({\bm
R})>,\nonumber\label{and}\\
\end{eqnarray}
where
\begin{eqnarray}
\eta&=&-i{\rm sgn}(\omega_m)\xi\sin\theta_ke^{i\phi_k}\sqrt{e\alpha\tilde{B}},\label{eta}
\end{eqnarray}
 the $L_n^0(x)$ are the Laguerre polynomials, and the matrix elements in Eq. (\ref{and}) are evaluated exactly as by L{\"o}rscher {\it et
 al.}\cite{Loerscher}, although the integrations over $f^2(\tilde{\bm k}')$ are more complicated than in that $p$-wave case with broken symmetry.

From Eq. (\ref{and}), we derived a recursion relation for the $a_{n,d}$ corresponding to the  $d_{x^2-y^2}$-wave order parameter,
\begin{eqnarray}
0&=&C_{n,n-4}a_{n-4,d}+C_{n,n-2}a_{n-2,d}+C_{n,n}a_{n,d}\nonumber\\
& &+C_{n,n+2}a_{n+2,d}+C_{n,n+4}a_{n+4,d},
\end{eqnarray}
where the coefficients $C_{n,n'}$ are given in the Appendix.

\section{Numerical Results}

\begin{figure*}
\begin{center}
\includegraphics[width=0.3\linewidth]{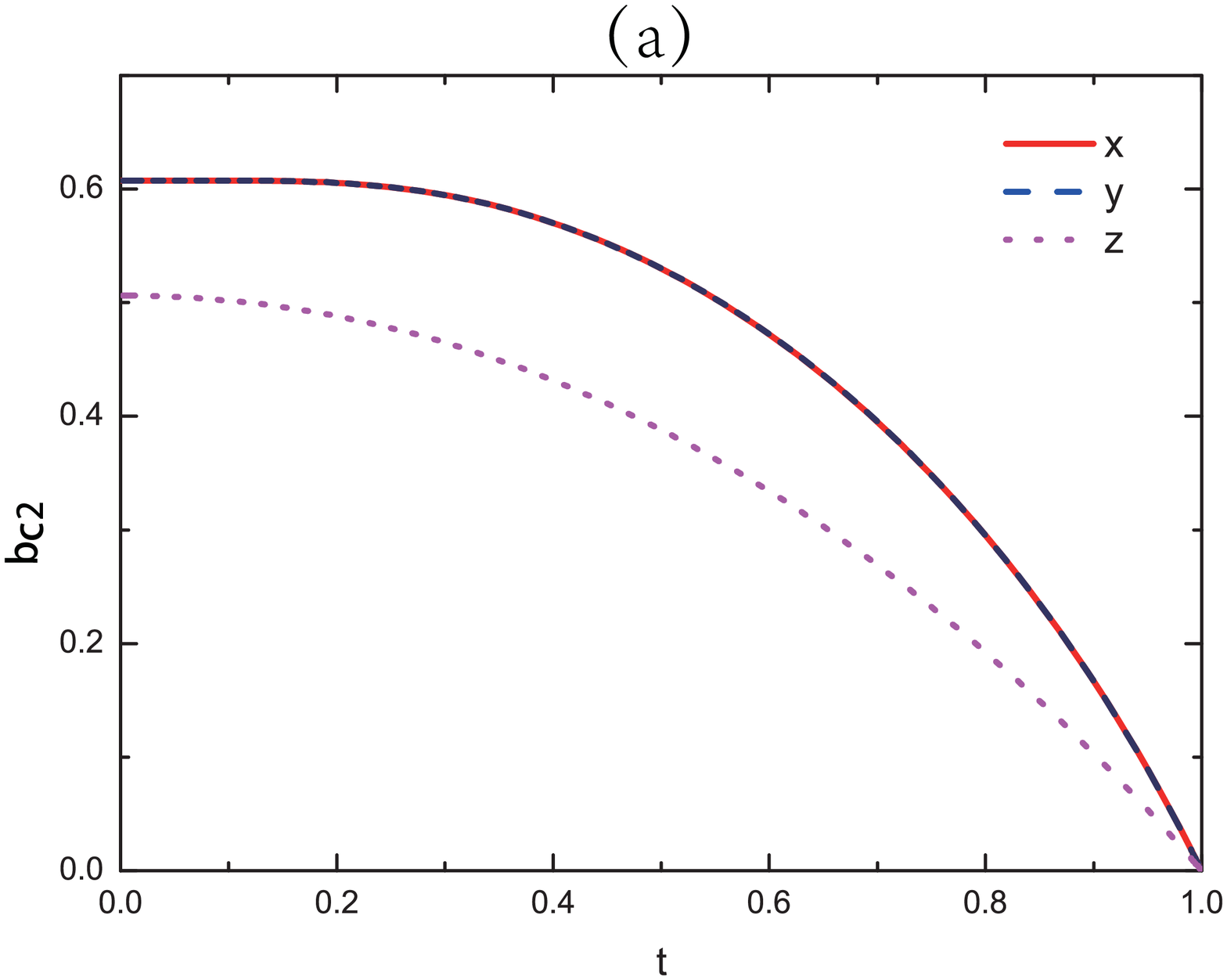}
\includegraphics[width=0.3\linewidth]{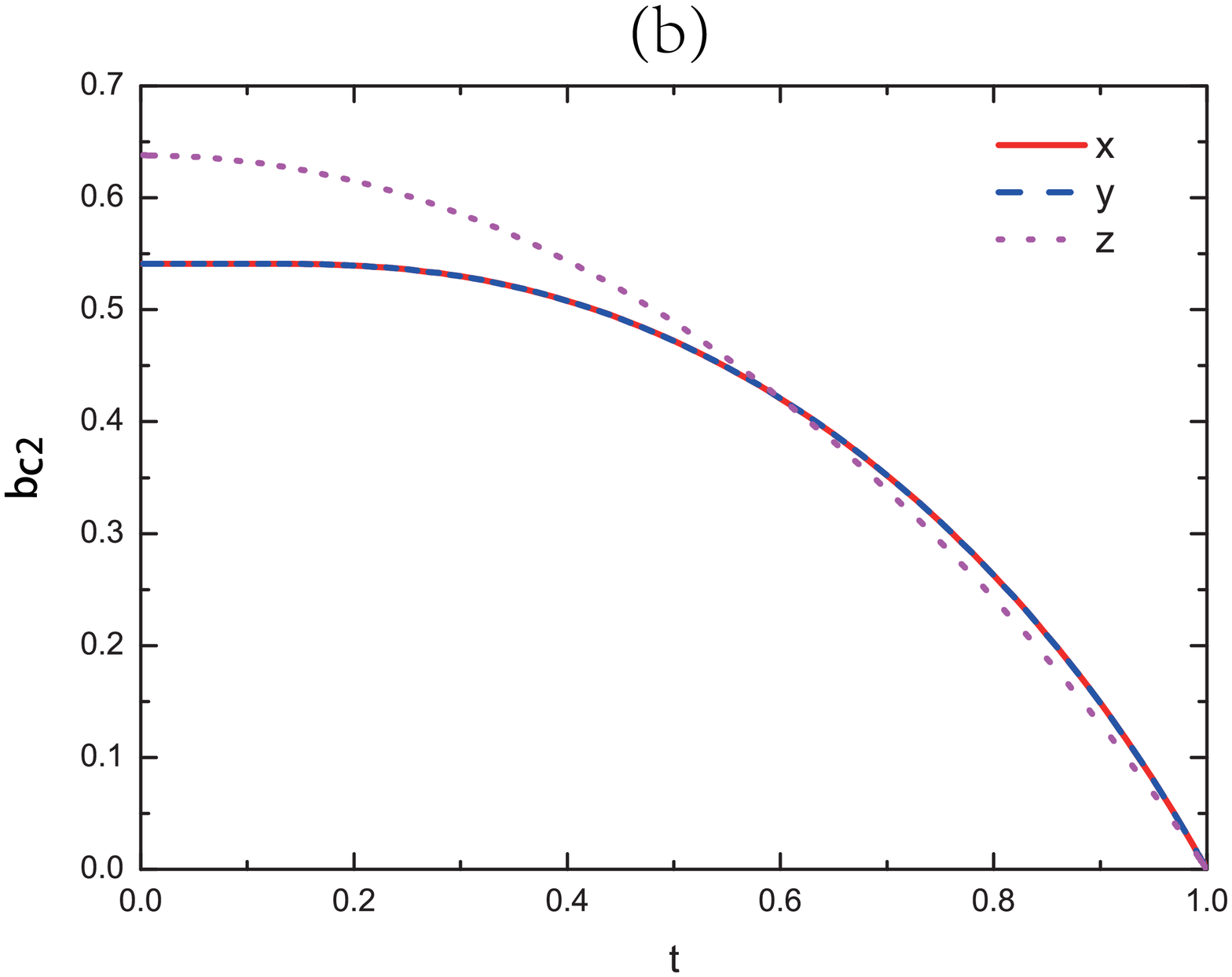}
\includegraphics[width=0.3\linewidth]{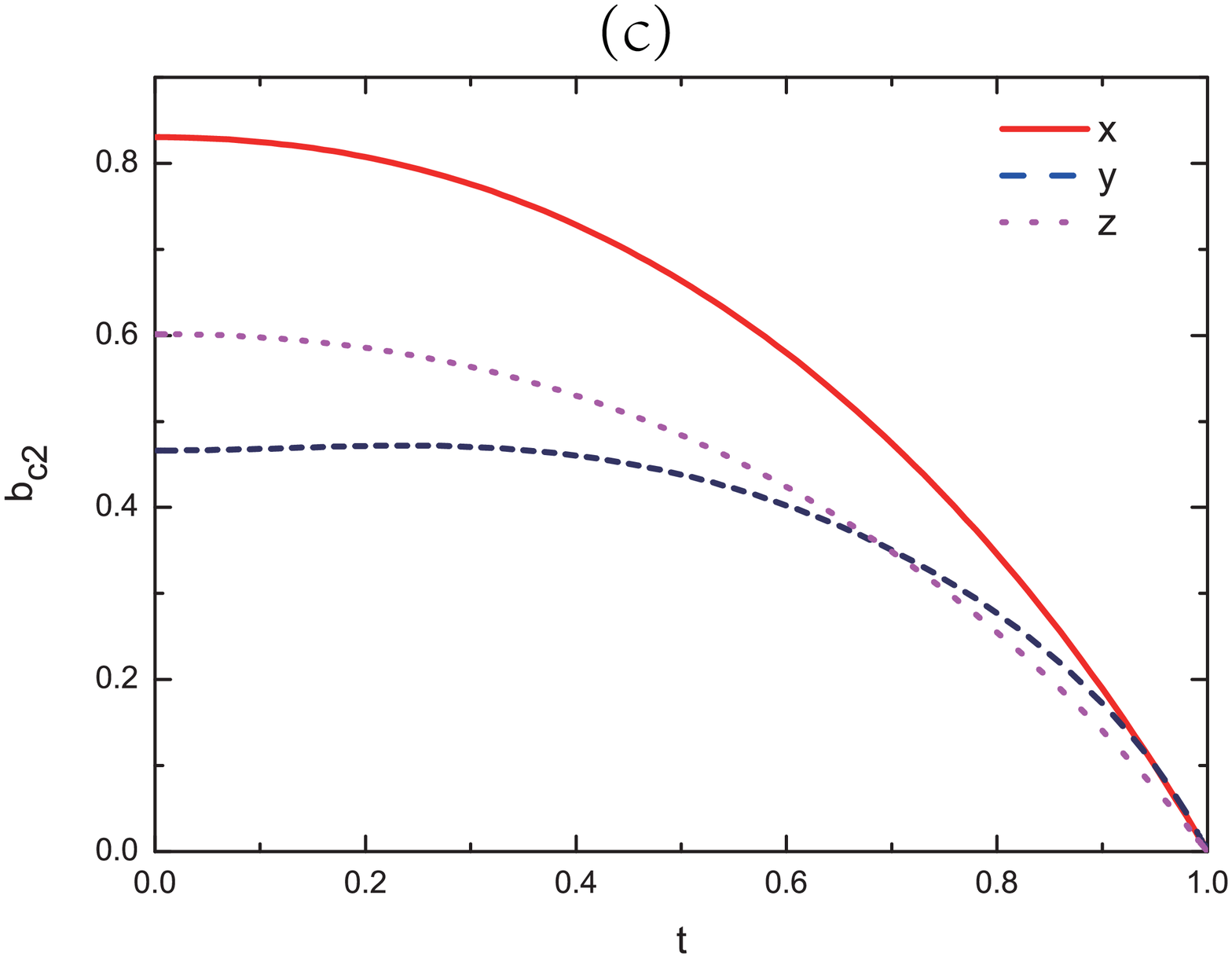}
\vspace{-5mm}
\end{center}
\caption{(Color online) The reduced upper critical field $b_{c2}(t)$  for a $d_{x^2-y^2}$-wave superconductor with the field along $\hat{\bm x}$ (solid
red), $\hat{\bm y}$ (dash blue), and $\hat{\bm z}$ (dot magenta) directions, all for which $\overline{m}_{1}\overline{m}_{2}\overline{m}_{3}=1$.
(a) $\overline{m}_1=\overline{m}_2=\overline{m}_3=1$. (b) $\overline{m}_1=\overline{m}_2=1.26$, $\overline{m}_3=0.63$. (c) $\overline{m}_1=0.5$,
$\overline{m}_2=2.0$, $\overline{m}_3=1.0$.}
\label{fig:3}
\end{figure*}

\begin{figure*}
\begin{center}
\includegraphics[width=0.3\linewidth]{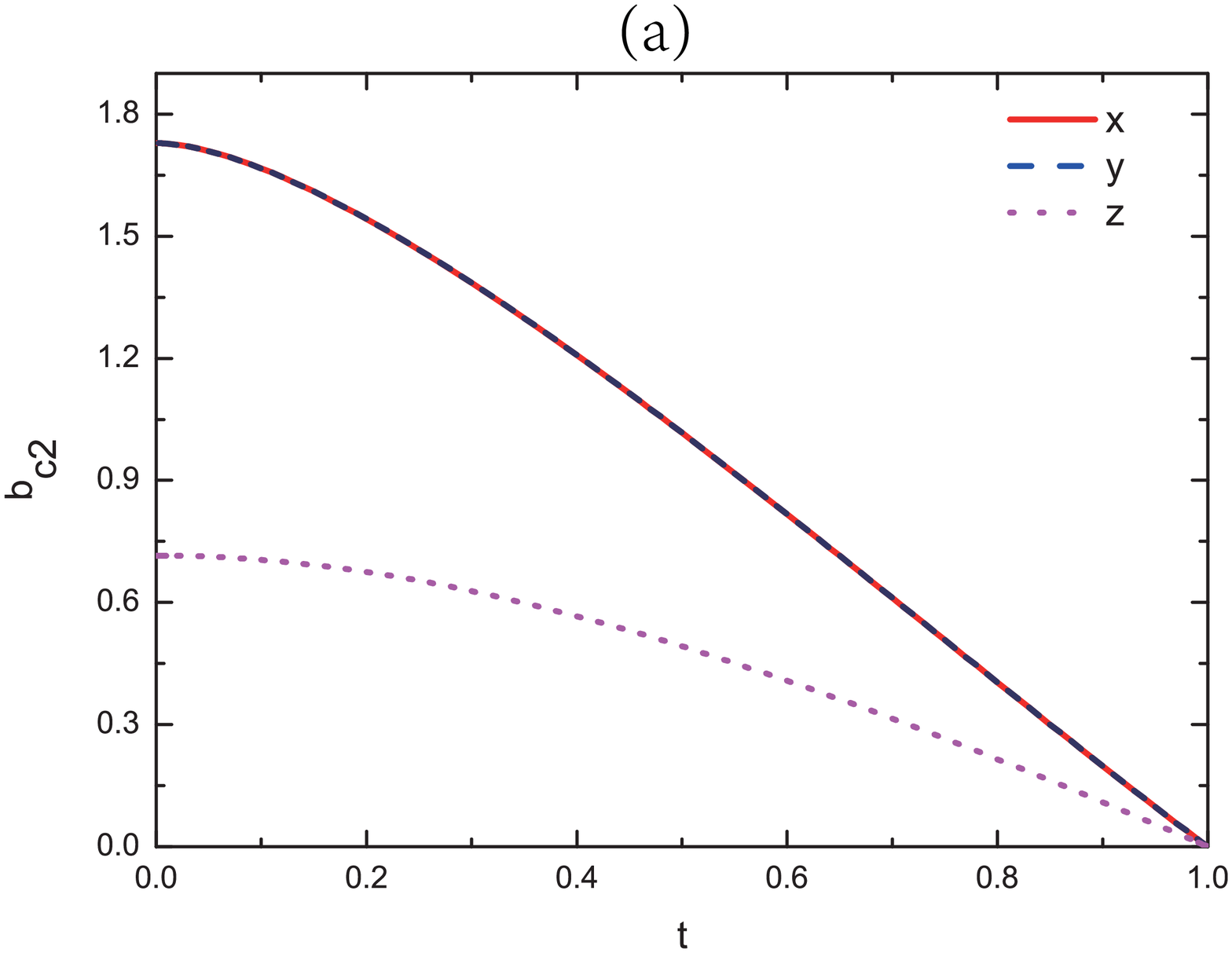}
\includegraphics[width=0.3\linewidth]{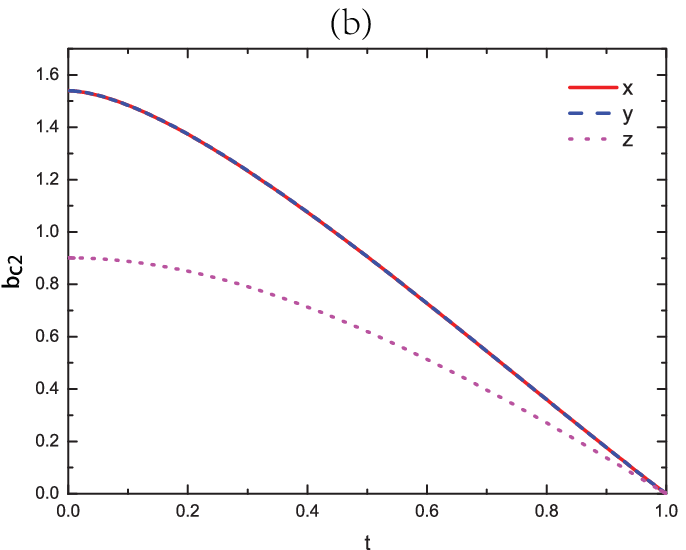}
\includegraphics[width=0.3\linewidth]{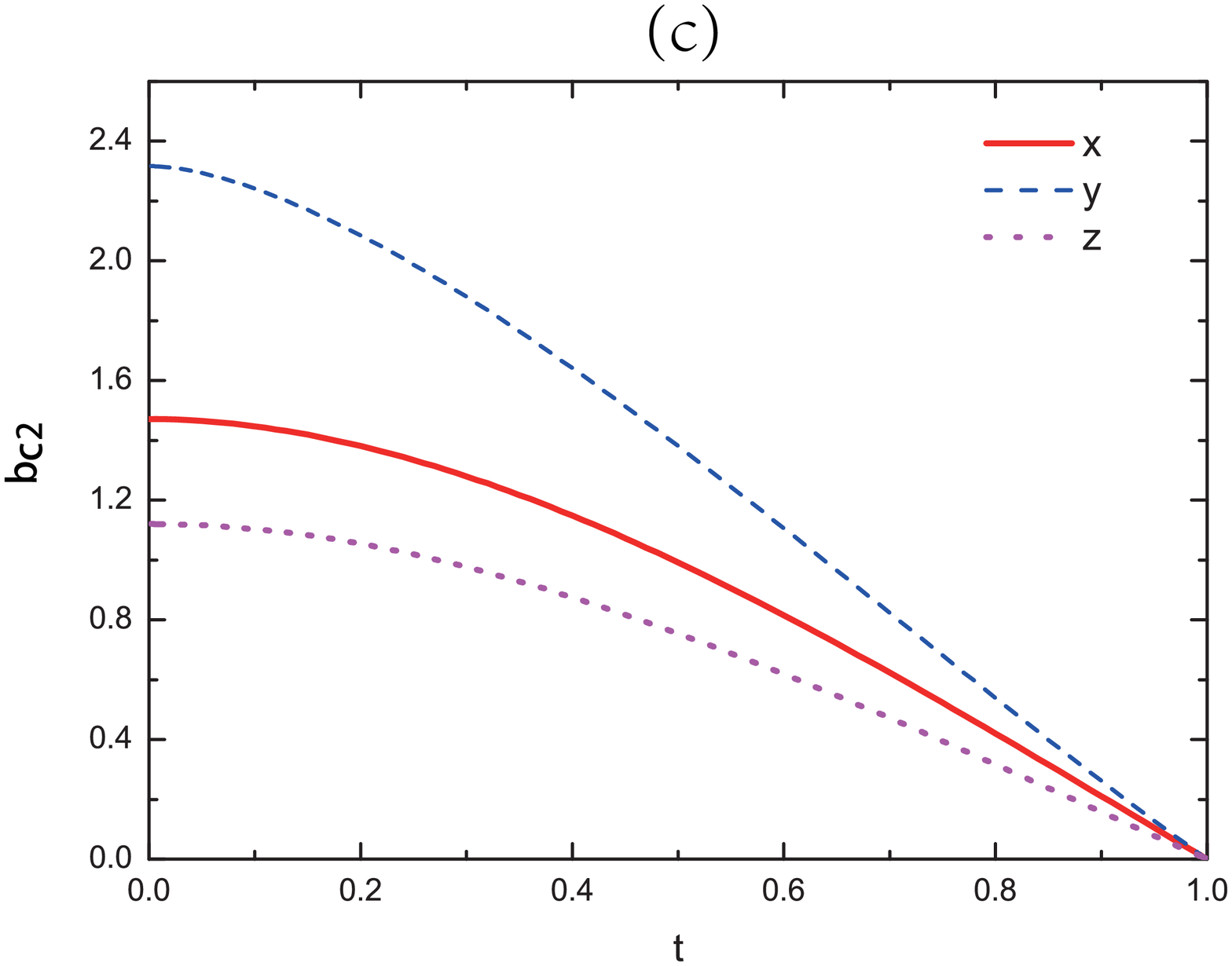}
\vspace{-5mm}
\end{center}
\caption{(Color online)  The reduced $b_{c2}(t)$ for $d_{x^2-y^2}$-wave superconductor without the Zeeman energy.  Figures (a), (b), and (c)
correspond to the same sets of $\overline{m}_i$ values as in Fig. 3, and the field directions are also indicated exactly as in Fig. 3.}
\label{fig:4}
\end{figure*}

Here we present our numerical results for the  reduced upper critical induction $b_{c2}$ and reduced temperature $t$ defined by
\begin{eqnarray}
b_{c2}(\theta,\phi,t)&=&2e\tilde{B}_{c2}(\theta,\phi,t)v_{F}^2/(2\pi T_{c})^2,\label{bc2}\nonumber\\
t&=&T/T_c
\end{eqnarray}
for $s$-wave and $d_{x^2-y^2}$-wave superconductors, and discuss the influences of the anisotropic effective mass, Zeeman energy, and the order parameter (OP) symmetry.

First, we present our results for $B_{c2}$ of an $s$-wave superconductor with an anisotropic effective mass and Zeeman tensor, but which has an isotropic
superconducting gap on the basis of Section \uppercase\expandafter{\romannumeral2}. In Fig. 2(a), $b_{c2}$ for an $s$-wave superconductor is plotted versus $t$ for a spherical FS and for a FS with ellipsoidal anisotropy. Clearly, $b_{c2}$ is independent of the direction of the $s$-wave OP on an isotropic FS, as shown  by the solid black line in Fig. 2(a). However, that is not the case for an $s$-wave OP on an anisotropic FS, and one example is presented in Fig. 2(a) for $\overline{m}_{1}=0.5, \overline{m}_{2}=2.0, \overline{m}_{3}=1.0$. The reduced upper critical field $b_{c2}$ depends upon the field direction, and $b_{c2}$ is largest in the direction corresponding to the smallest effective mass. Actually, the difference arises from the first step of Klemm-Clem transformation, in which ${\bm B}^{\prime}(\theta^{\prime},\phi^{\prime})$ is $\alpha(\theta,\phi)$ times the magnitude of ${\bm B}(\theta,\phi)$, but the gap function is a constant in wave vector space. So it is clear that when the FS is isotropic, $\alpha=1$, and $b_{c2}$ for an isotropic $s$-wave superconductor is independent of the field direction.

We note that the $b_{c2}$ as shown in Fig. 2(a) is  larger in the direction with of the smaller effective mass.  We therefore investigated the effects of the much stronger effective mass anisotropy but also  with an
isotropic $s$-wave pairing interaction. In Figs. 2(b) and 2(c), The $b_{c2}$ for the fields along the $x$, $y$, and $z$ directions are
depicted respectively in solid red, dashed blue, and dotted black  curves. In Fig. 2(b), the reduced effective mass values are $\overline{m}_1=10^{-4}$,
$\overline{m}_2=1.0$, and $\overline{m}_3=10^4$, and $\overline{m}_1=\overline{m}_2=0.005$, $\overline{m}_3=4.0\times10^4$ in Fig. 2(c). For the field along the largest effective mass direction, $b_{c2}(t)$ is so small that we displayed the results in the figure insets.

 Helfand and Werthamer\cite{Helfand1966} investigated the upper critical field for an $s$-wave superconductor on an isotropic FS in then clean limit, and found that its slope at $T_c$ was given by $B_{c2}=0.73|d H_{c2}/d T|_{T=T_c} T_c$. We also obtained the same value, for which the slope for the reduced upper critical field $b_{c2}$ at $t=1$ is $-1.426$ for an  isotropic FS.  We  also found that the slope of $B_{c2}(t)$ at $T_c$ for an  anisotropic FS is $B_{c2}(t)=\frac{b_{c2} \alpha}{12/7 \zeta(3)}|d B_{c2}/d T|_{T=T_c} T_c $. There is a chance that if this slope were to persist to low temperatures, $B_{c2}(0)$  could exceed the conventional Pauli limiting $B_p ({\rm T})=1.76 T_c ({\rm K})$, obtained without any effective mass anisotropy, provided that $\frac{b_{c2} \alpha}{12/7 \zeta(3)} $ is greater than $1.76$, where $\alpha$ is given by Eq. (\ref{alpha}).

\begin{figure*}
\begin{center}
\includegraphics[width=0.3\linewidth]{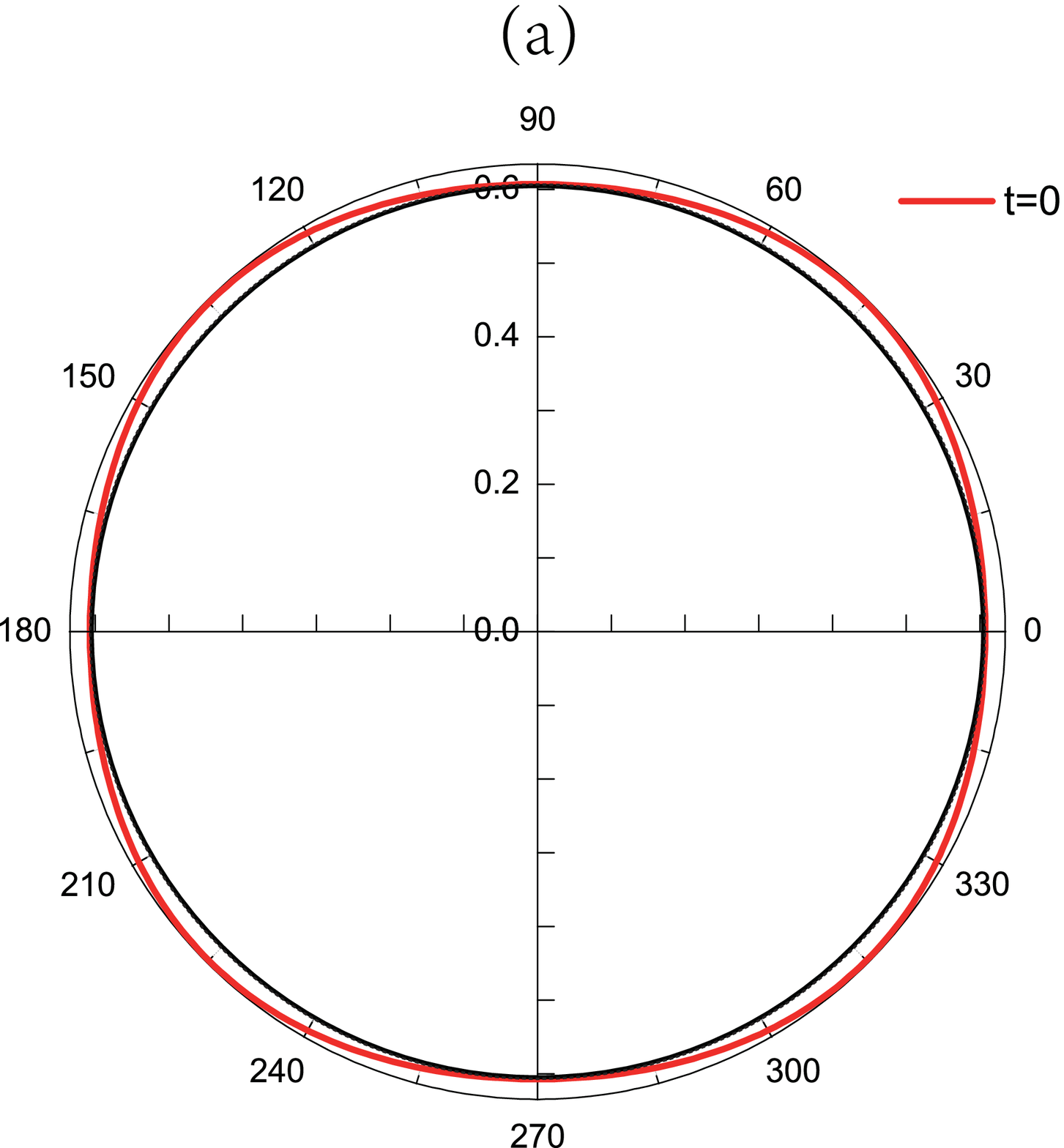}
\includegraphics[width=0.3\linewidth]{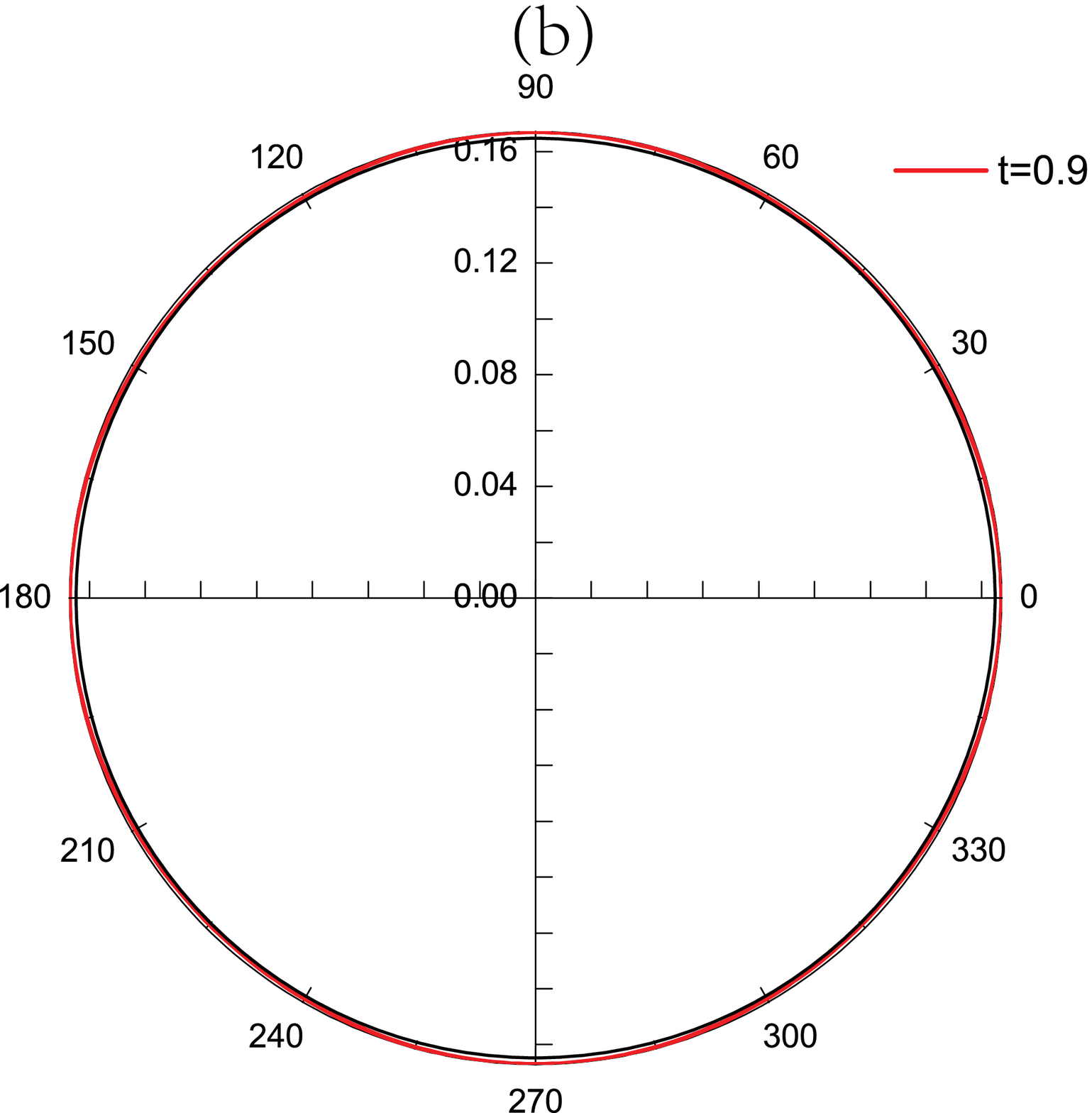}
\includegraphics[width=0.3\linewidth]{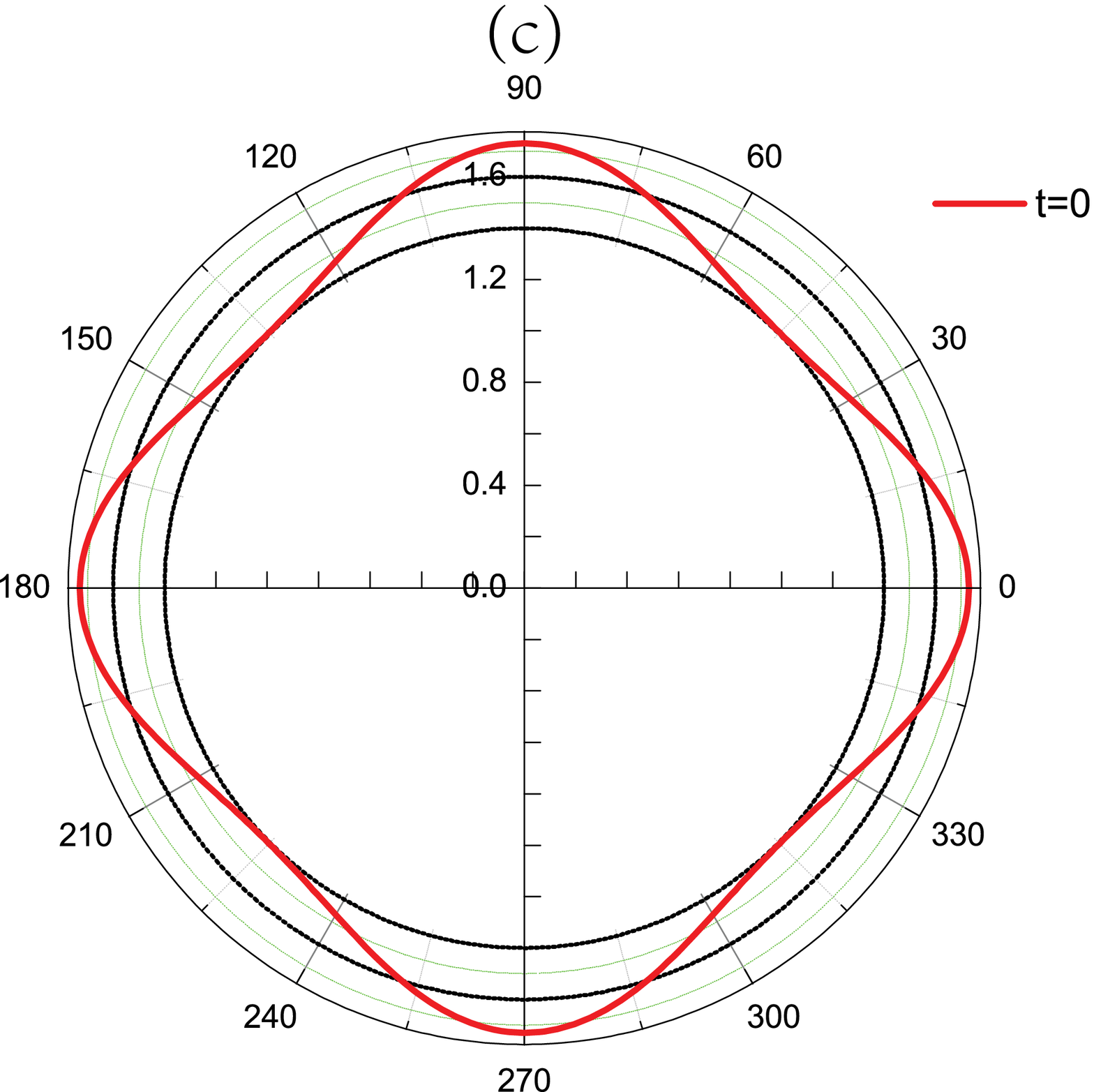}
\vspace{-5mm}
\end{center}
\caption{(Color online) In-plane $b_{c2}(\pi/2,\phi,t)$ for a $d_{x^2-y^2}$-wave superconductor. (a) and (b) show $b_{c2}(\pi/2,\phi,t)$ for a $d_{x^2-y^2}$-wave superconductor in the $xy$-plane including the Zeeman energy for an isotropic FS at $t=0$, and $t=0.9$, respectively. (c) shows the results for $b_{c2}(\pi/2,\phi,0)$ for a $d_{x^2-y^2}$-wave superconductor with an isotropic FS but without the Zeeman energy at $t=0$.}
\label{fig:5}
\end{figure*}

\begin{figure*}
\begin{center}
\includegraphics[width=0.24\linewidth]{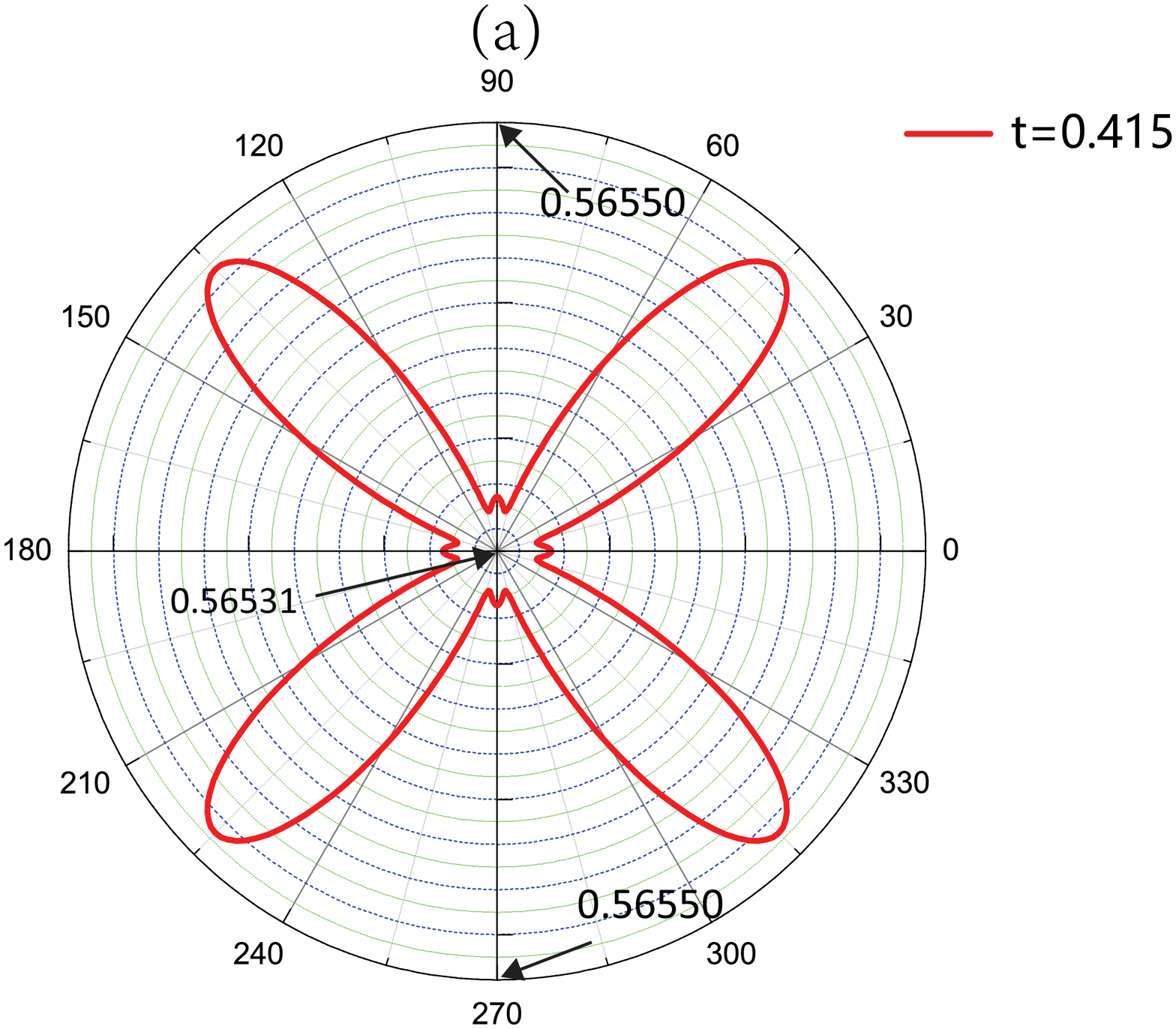}
\includegraphics[width=0.24\linewidth]{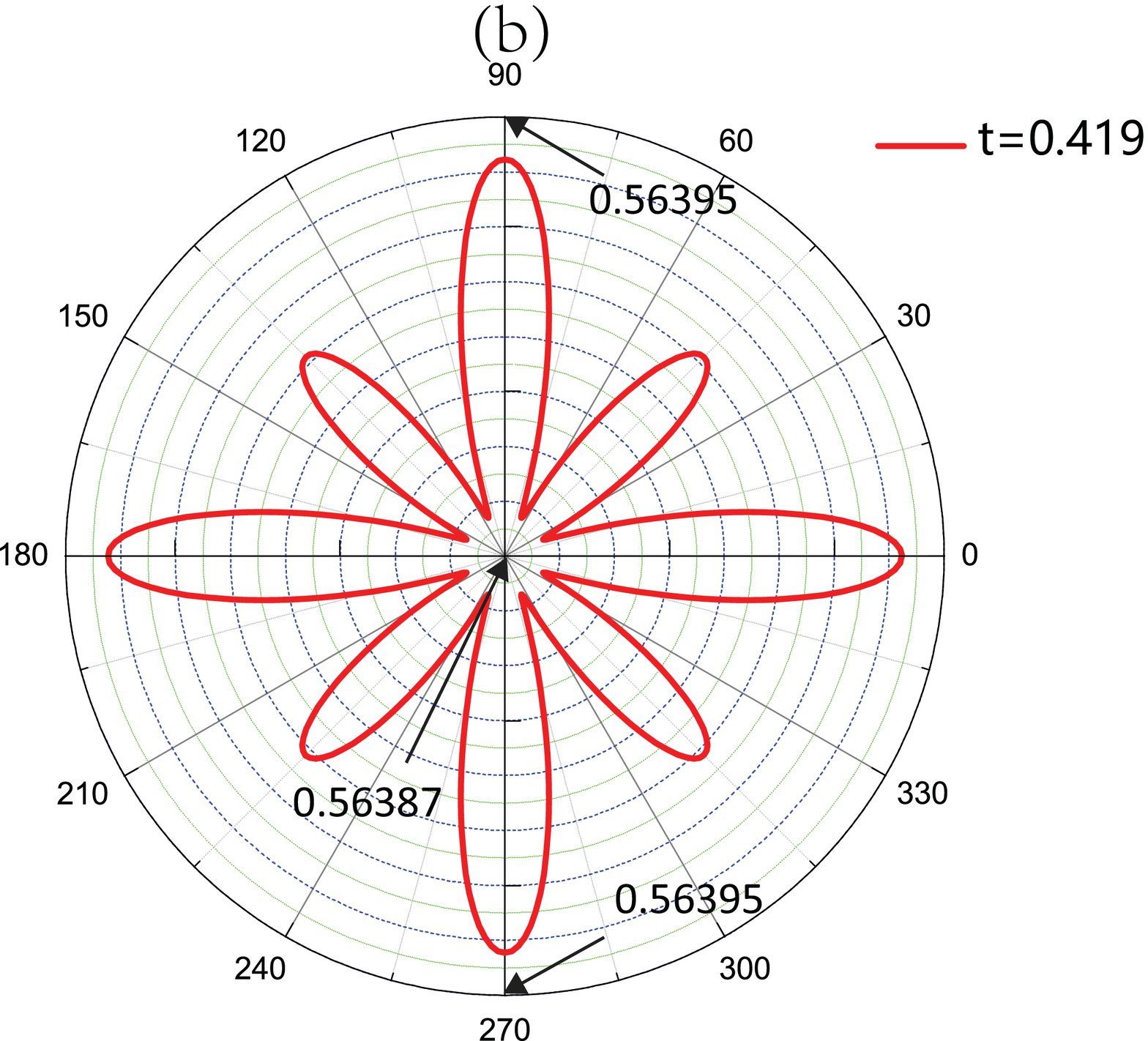}
\includegraphics[width=0.24\linewidth]{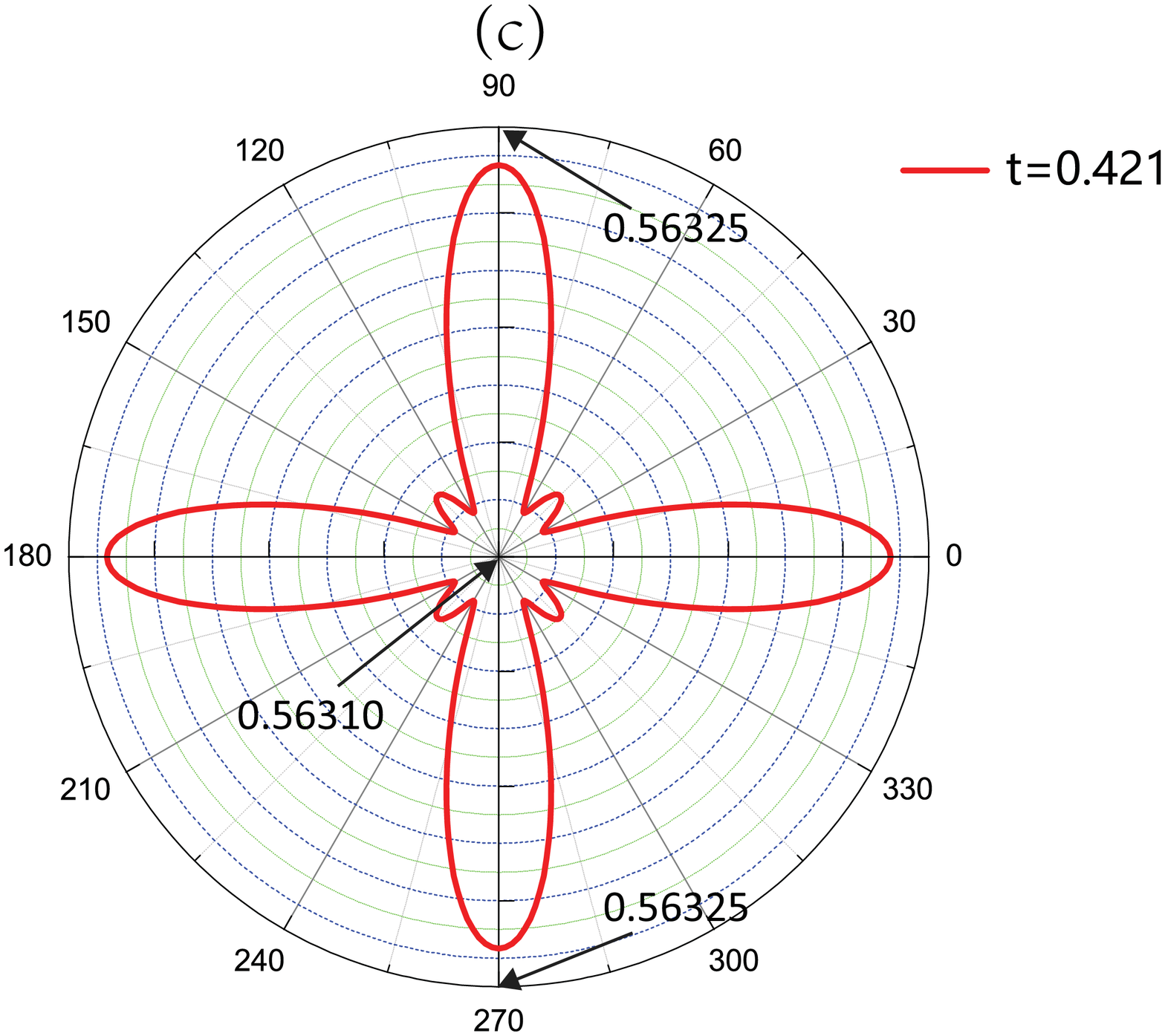}
\includegraphics[width=0.24\linewidth]{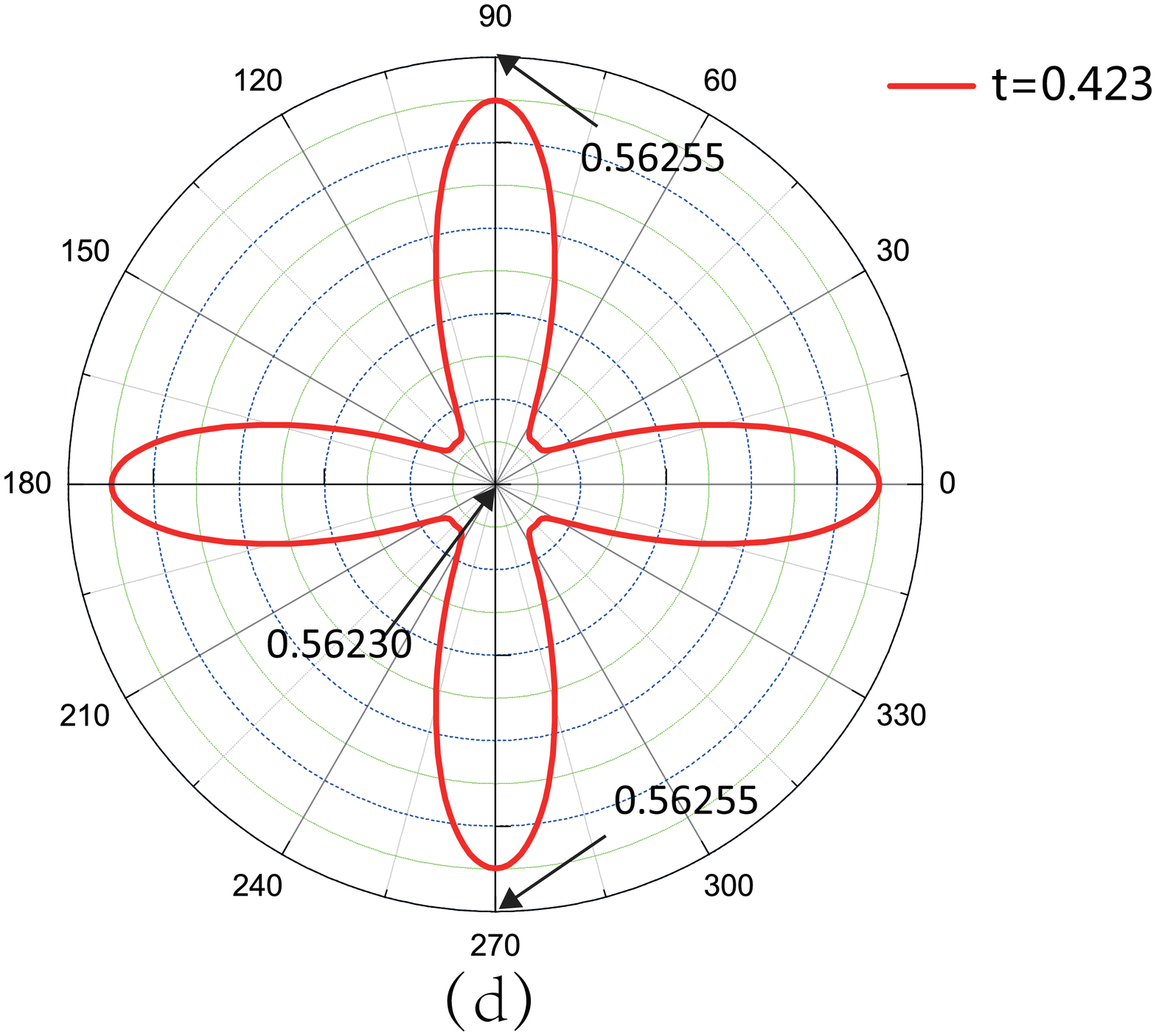}
\vspace{-5mm}
\end{center}
\caption{(Color online) Here we present some details of the region of rapid change in the symmetry of $b_{c2}(\pi/2,\phi,t)$ for a $d_{x^2-y^2}$-wave
order parameter with an isotropic FS without the Zeeman interaction.  From left to right, $t=0.415, 0.419, 0.421$ and $0.423$.}
\label{fig:6}
\end{figure*}

\begin{figure*}
\begin{center}
\includegraphics[width=0.32\textwidth]{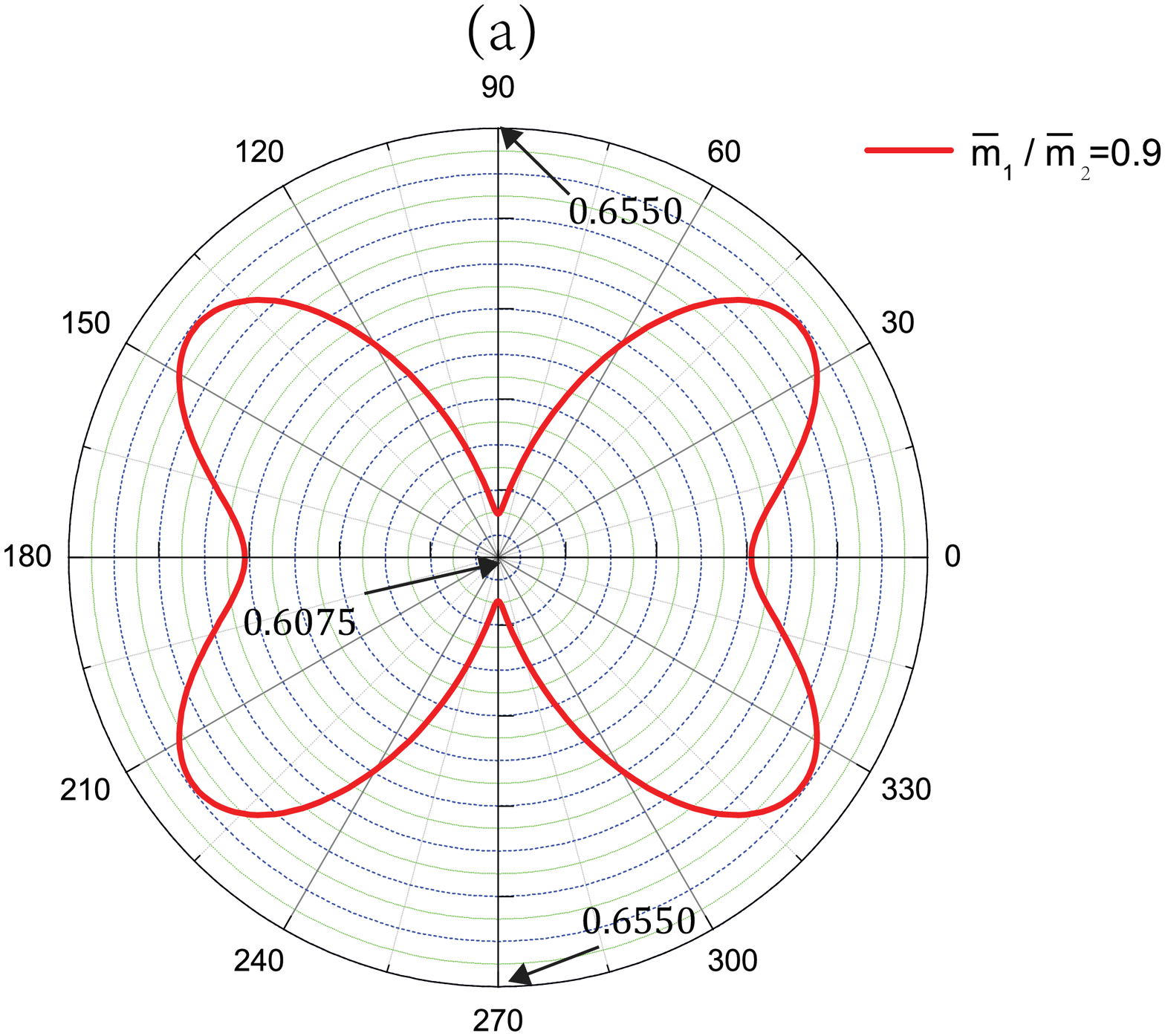}
\includegraphics[width=0.32\textwidth]{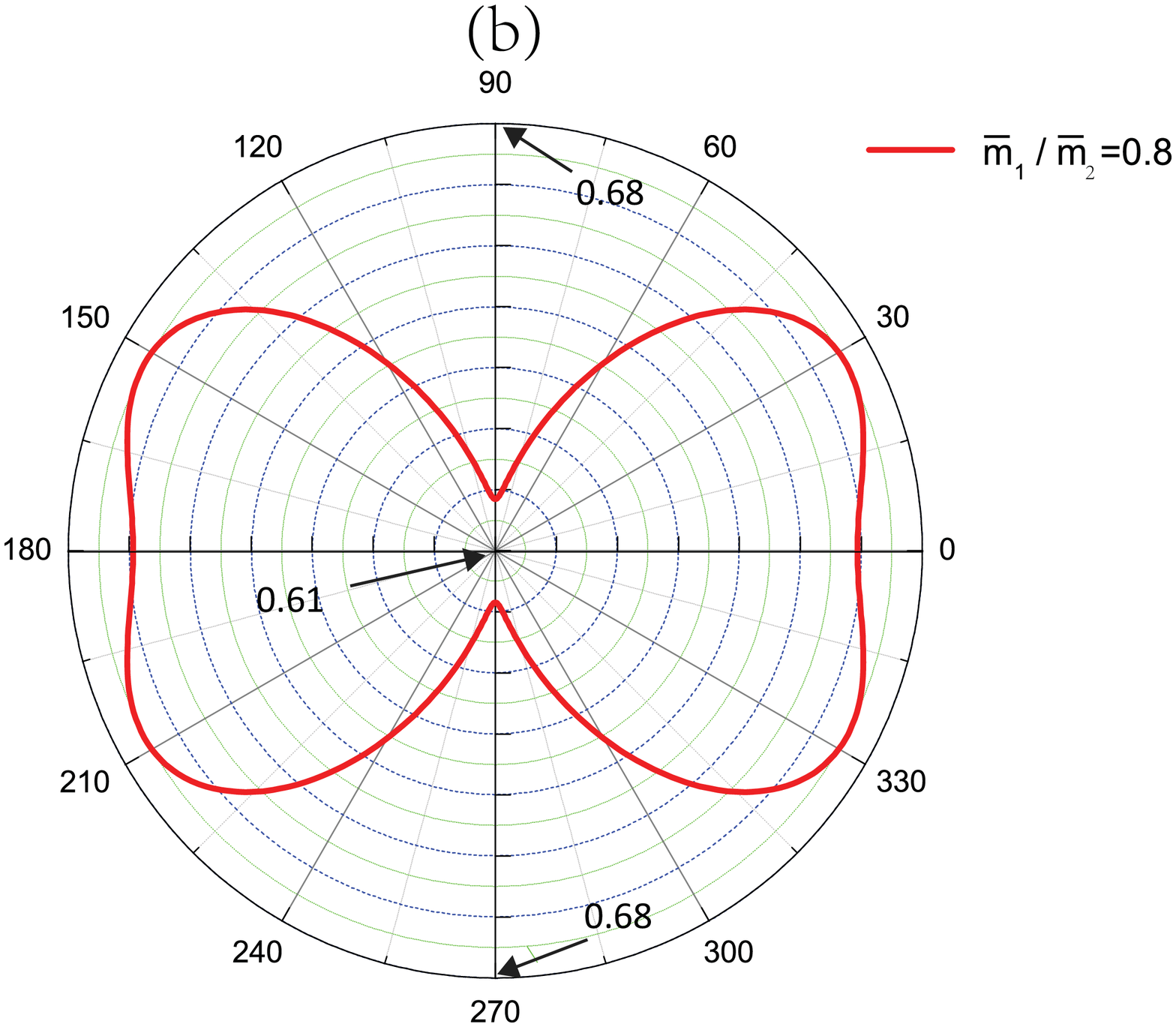}
\includegraphics[width=0.32\textwidth]{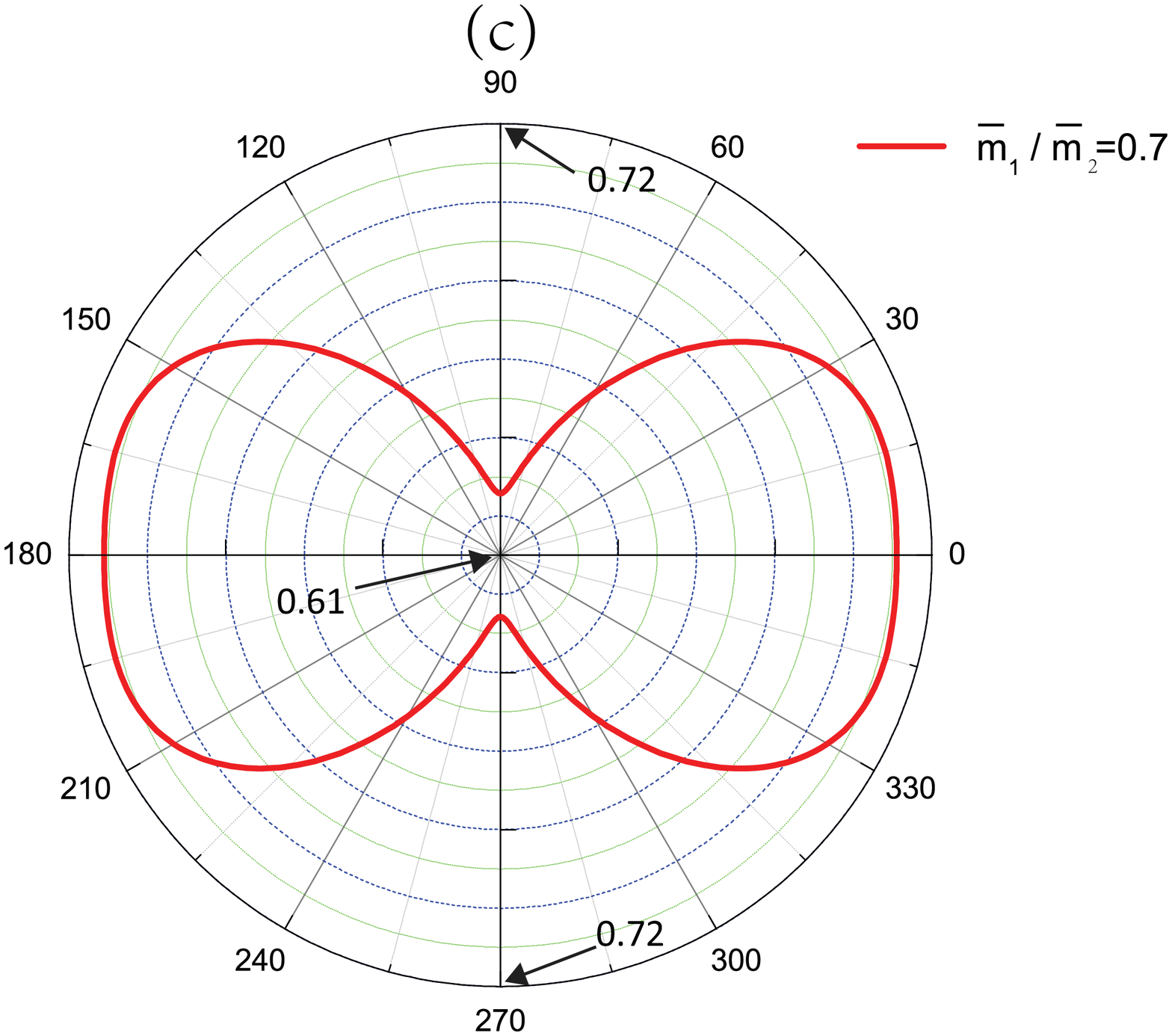}
\vspace{-5mm}
\end{center}
 \caption{(Color online) Azimuthal dependence at $t=0$ of $b_{c2}(\pi/2,\phi,0)$ for a $d_{x^2-y^2}$-wave superconductor with decreasing
$\overline{m}_1/\overline{m}_2$ at fixed $\overline{m}_1 \overline{m}_2 \overline{m}_3=1$ and $\overline{m}_2=1$. From left to right:  $\overline{m}_1=0.9$, 0.8, and 0.7.  In all three of these
figures, $b_{c2}(\pi/2,\phi,0)$ has a dumbbell shape exhibiting $C_2$ symmetry, but the directions of the maxima rotate from the $x$ axis for
$\overline{m}_1=0.7$ to increasingly off that axis as $\overline{m}_1$ increases.}
\label{fig:7}
\end{figure*}

\begin{figure*}
\begin{center}
\includegraphics[width=0.32\linewidth]{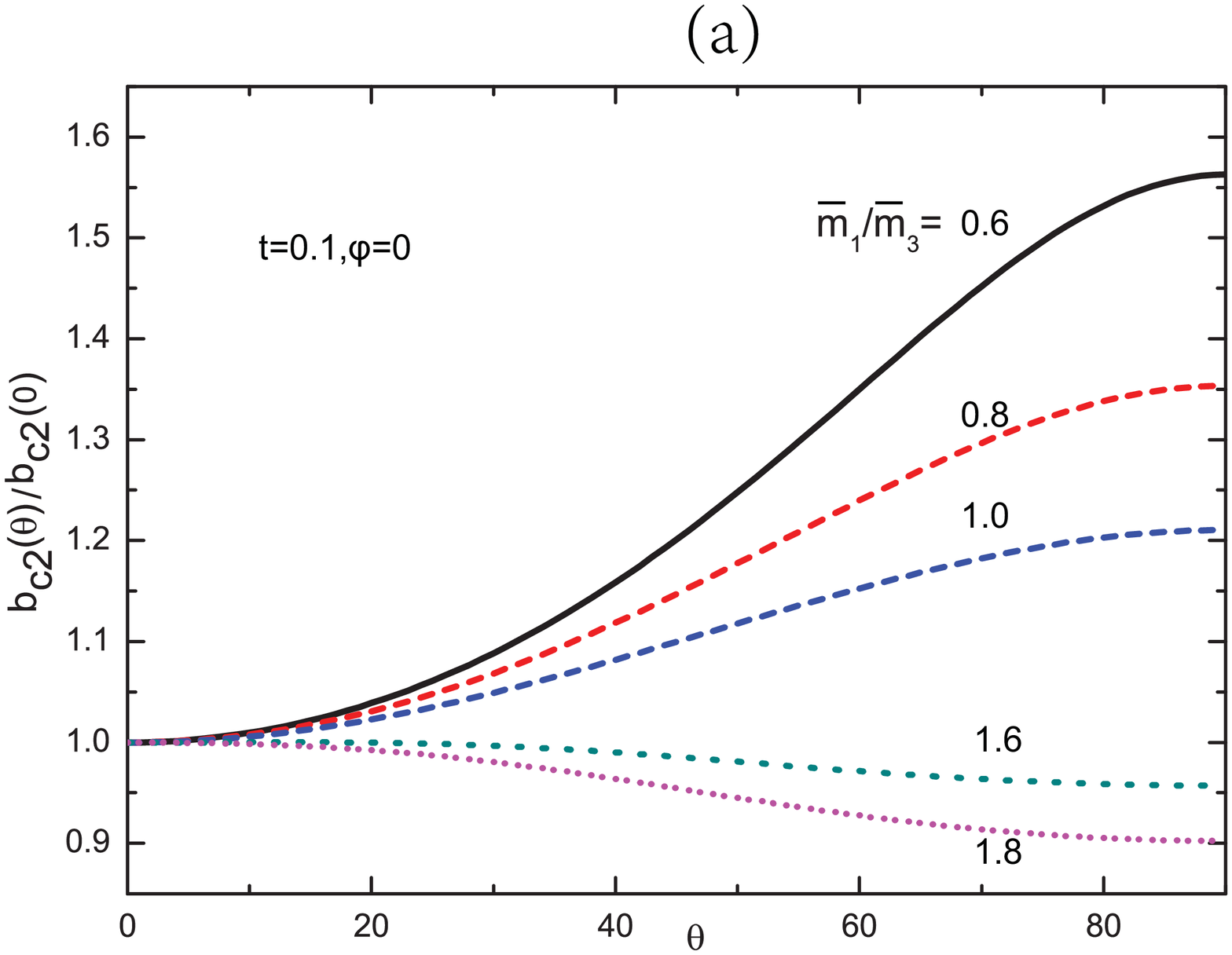}
\includegraphics[width=0.32\linewidth]{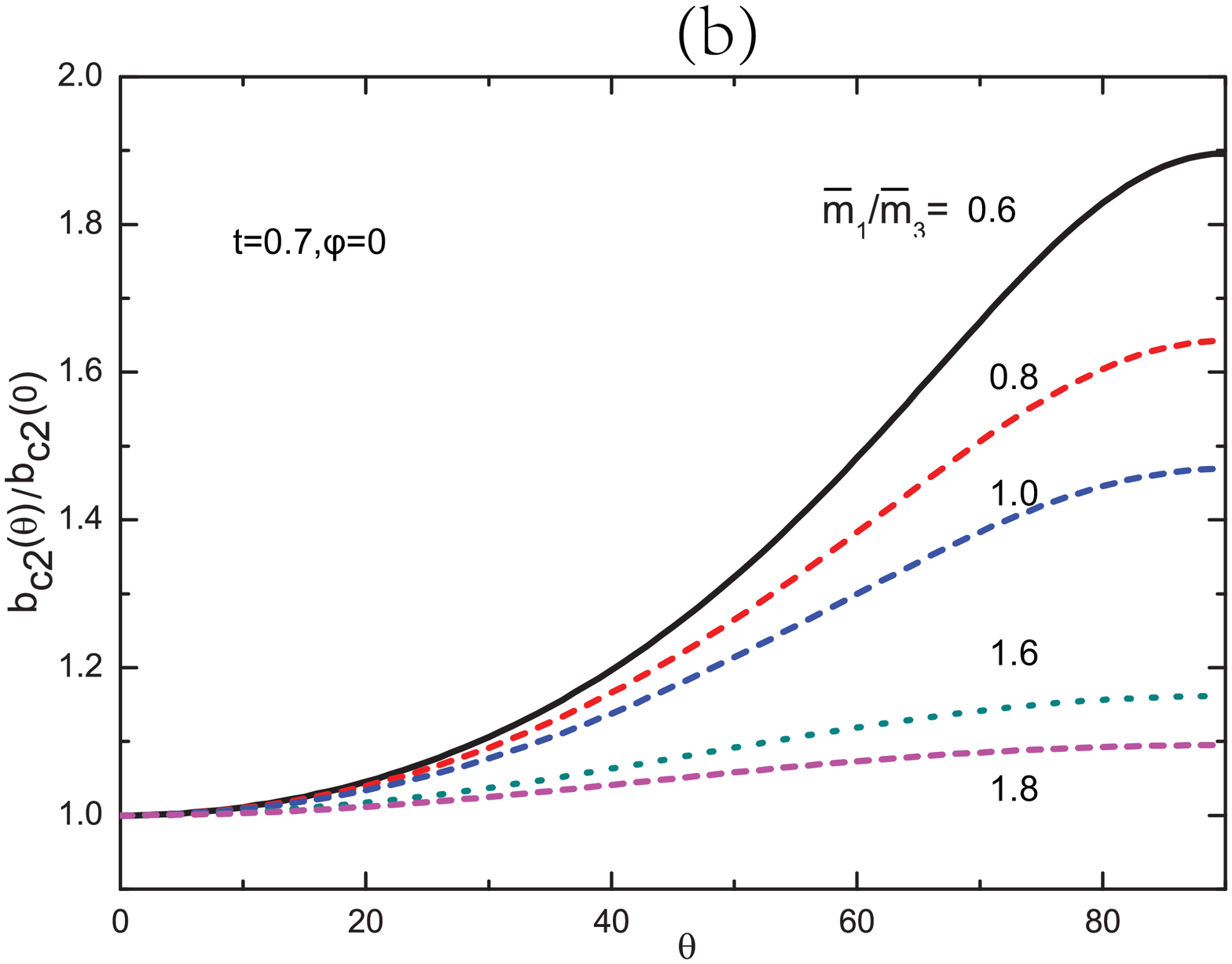}
\includegraphics[width=0.32\linewidth]{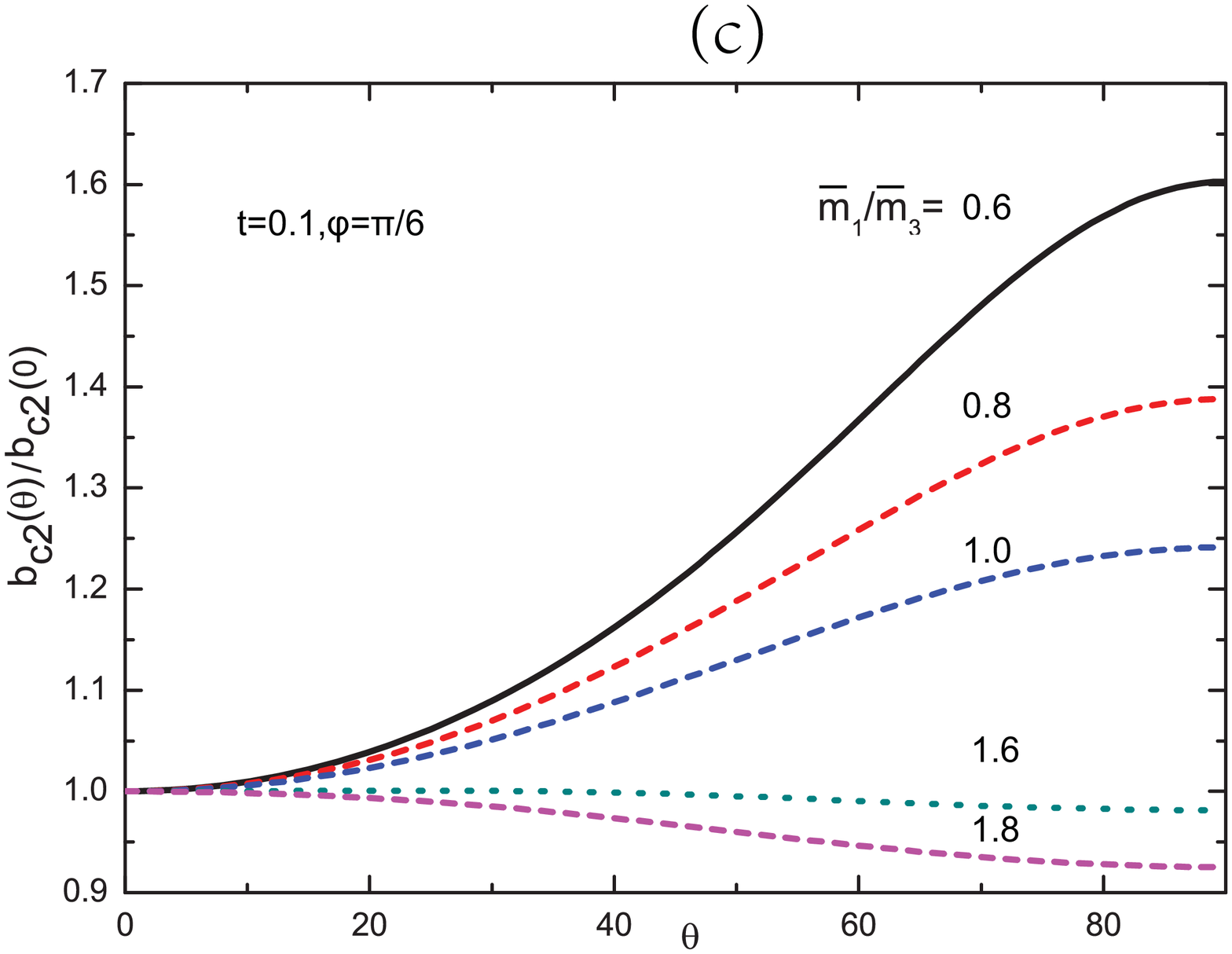}
\vspace{-5mm}
\end{center}
 \caption{(Color online) Shown are plots of the polar ($\theta$) angular dependence of  $b_{c2}(\theta,\phi,t)/b_{c2}(0,0,0)$ normalized to its
$\theta=0^{\circ}$ value for a $d_{x^2-y^2}$-wave superconductor as $t\rightarrow0$, for $\overline{m}_1=\overline{m}_2$,
and $\overline{m}_1/\overline{m}_3 =0.6, 0.8, 1.0, 1.6$ and $1.8$ respectively. For each tetragonal effective mass anisotropy pictured, $b_{c2}(\theta,\phi,t)$ is a
monotonically increasing or decreasing  function of $\theta$ at fixed $t$ and $\phi$.}
\label{fig:8}
\end{figure*}

However, the situation is considerably more complex for a $d_{x^{2}-y^{2}}$-wave OP than for an $s$-wave OP, especially for an anisotropic FS. Firstly, after the transformations, the  interaction and the
$d_{x^{2}-y^{2}}$ OP both depend upon the directions in the pairing plane, as can be seen in Eqs. (17) and (20). In Fig. 3, we exhibit $b_{c2}(t)$ for a $d_{x^{2}-y^{2}}$-wave superconductor with the field along the $x$, $y$, and $z$ directions for three different cases, each with $\overline{m}_1\overline{m}_2\overline{m}_3=1$. In Fig. 3(a), the FS is isotropic, $b_{c2}(t)$ is identical for the field along the $x$- and $y$-axes,  and is  larger than for the field along the $z$-axis. The strong anisotropy in slope just below $t=1$ is solely due to the OP anisotropy, as the FS is isotropic. In fact, $b_{c2}$ for a  $d_{x^2-y^2}$-wave superconductor with an anisotropic effective mass becomes more complicated and difficult to describe, as shown in Figs. 3 (b) and (c), where $b_{c2}(\theta,\phi)(t)$ crossing points exist. In Fig. 3(b), $b_{c2}(t)$ along the $x$ and $y$ directions are identical, due to the equal effective masses in the $xy$ plane, but cross $b_{c2}(t)$ along the $z$ direction.  In Fig. 3(c), effective mass anisotropies on the order 2 can cause dramatically different $b_{c2}(t)$ curves for the three orthorhombic directions. We note that these curves include the relativistically consistent anisotropic Zeeman energy, and the slopes at $T_c$ are sufficiently large as to approach the conventional Pauli limit.

In Fig. 4, we show the $b_{c2}(t)$ curves for the $d_{x^2-y^2}$-wave superconductor with the same FS anisotropy parameters, but without the self-consistent treatment of the Zeeman interaction, as has often been incorrectly assumed in the literature.  In these figures, $g=0$ for all field directions, which is inconsistent with the anisotropic Dirac equation used to generate the non-relativistic limit of the Hamiltonian \cite{Aiying2021}.  In these figures, there are no crossing points of $b_{c2}(t)$ for the different field directions.
Further study shows that the such crossing points seem to disappear comparing with the results in Figs. 4 (b) (c) without the Zeeman energy, but to strengthen this argument requires further study. We only show a few examples in Figs. 4(b) and (c), for which the crossing points vanish. The crossing point will exist for a pancake FS  ($\bar{m}_1=\bar{m}_{2}\textgreater\bar{m}_3$) with the ratio $m_{1}/m_{3}$ between $1.6$ and $3.2$. Therefore, the Zeeman energy, which is intrinsically anisotropic, plays a crucial role in determining $b_{c2}(\pi/2,\phi,t)$ for a $d_{x^2-y^2}$-wave superconductor. At least, it suppresses the $B_{c2}(t)$ for $d_{x^2-y2}$-wave superconductor more strongly than for an $s$-wave superconductor. But there remains the  possibility that the conventional Pauli limit could be exceeded in some field directions for a $d_{x^2-y^2}$-wave superconductor with a highly anisotropic FS.

It is interesting to investigate $b_{c2}(\pi/2,\phi,t)$ in the $xy$ plane. Not only have many researchers recently paid more attention to measuring
$b_{c2}$ in that plane\cite{Zuo2017,Vorontsov2010}, many superconductors, such as the cuprates, have been thought by many authors to exhibit $d_{x^2-y^2}$ OP
symmetry\cite{Shen2003,Tsuei2000}. Here we present results for $b_{c2}(\pi/2,\phi,t)$ in the $xy$ plane, with and without the Zeeman interaction. It is intriguing to find that
$b_{c2}(\pi/2,\phi,t)$ including the Zeeman energy at different $t$ values  does not always follow the commonly held belief that the maxima $b_{c2}(\pi/2,\phi,t)$
always occurs along the antinodal directions of the of the OP. Instead, there is a $\pi/4$ azimuthal shift in these maxima if this slope were to persist to  low $t$.   That is, the maxima in
$b_{c2}(\pi/2,\phi,t)$ for a $d_{x^2-y^2}$-wave superconductor just below $t=1$ is indeed along the antinodal  directions, but at low $t$,  the maxima are along the nodal directions.

This change in symmetry is shown
in Figs. 5(a) and (b), which are the results for an isotropic FS at $t = 0$ and  $t = 0.9$, respectively. Fig. 5(c) presents the results for $b_{c2}(\pi/2,\phi,t)$
without the Zeeman energy in the  $xy$ plane at low $t$. The shape of $b_{c2}(\pi/2,\phi,t)$ in Fig. 5(c) without the Zeeman energy actually is similar to
that of Fig. 5 (b), but the four-fold anisotropy is greatly reduced in magnitude by the Zeeman energy in Fig. 5(b). This is easy to understand from Eq.
(21). The Zeeman energy suppresses the upper critical field for a $d_{x^2-y^2}$-wave superconductor and enters into the gap equation in the $\cos[(\sigma-\sigma^\prime)\frac{e}{2 m_g} \tilde{B}\xi]$ term, which can be expanded in powers of $\tilde{\bm B}$, yielding the factor $ 1-[(\sigma-\sigma^\prime)\frac{e}{2 m_g}\alpha\tilde{B}\xi]^2/2$, which is nearly unity near to $t=1$. However, at large $B_{c2}$ at low $t$, the $\pi/4$ phase shift demonstrates that the Zeeman energy is very important at low $t$.

Because $b_{c2}(\pi/2,\phi,t)$  for a $d_{x^2-y^2}$-wave superconductor with an isotropic FS with or without the Zeeman interaction is greatly different  at $t=0$ and $t=0.9$, we explored the process of the $\pi/4$
phase change in its $\phi$-dependence, and found that it changes rapidly  in the mid-$t$ range, as shown in Fig. 6 without the Zeeman interaction to emphasize the changes in the azimuthal dependence of $b{c2}(\pi/2,\phi,t)$ with decreasing $t$. As $t$ is lowered from 0.9,
$b_{c2}(\pi/2,\phi,t)$ initially increases very slightly along the antinodal directions down to 0.421, and then the intensity along the nodal directions
starts to grow quickly, finally taking over around $t=0.415$.

We also considered the effects of different effective masses on $b_{c2}(\pi/2,\phi,t)$ for a $d_{x^2-y^2}$-wave superconductor at low $t$. When the
effective mass anisotropy within the pairing $xy$ plane becomes strong, as in detwinned orthorhombic YBa$_2$Cu$_3$O$_{7-\delta}$, the in-plane azimuthal
angular dependence of $b_{c2}$ not only deviates strongly from the $C_4$ symmetry that the nominal $d_{x^2-y^2}$ Cooper pair obeys if the crystal were to
be tetragonal, its shape changes from the four-leaf clover of $C_4$ symmetry  to that of a dumbbell with $C_2$ symmetry. The precise form of the in-plane azimuthal anisotropy of $b_{c2}(\pi/2,\phi,0)$ for an orthorhombic $d_{x^2-y^2}$-wave
superconductor changing with the $\overline{m}_1/\overline{m}_2$ effective mass anisotropy is pictured in Fig. 7. In this figure, we show the results of our calculations for a fixed geometric mean
$\overline{m}_1\overline{m}_2\overline{m}_3=1$ and $\overline{m}_2=1$, with $\overline{m}_1=0.9$, 0.8, and 0.7 in Figs. 7(a), (b), and (c), respectively.  For the least in-plane anisotropy pictured in Fig. 7(a), the predicted $\pi/4$ rotation of the maxima in the four-fold azimuthal anisotropy shown in Figs. 5(a) and 6(a) is partly present for two of the maxima, but the figure shown in Fig. 7(a) has overall $C_2$ symmetry.  At $\overline{m}_1/\overline{m}_2=0.7$, the azimuthal anisotropy of $b_{c2}(\pi/2,\phi,0)$ has rotated to the $x$ axis, and has the simpler dumbbell shape.

A peak in the polar dependence of $b_{c2}$ was predicted for $p$-wave superconductors with completely broken symmetry for anisotropies corresponding to
$\overline{m}_3/\alpha^2(\pi/2,\phi)>3$\cite{Loerscher}, we tried to study whether a peak in $b_{c2}(\theta, \phi,t)$ at some fixed $\phi$ and $t$ might occur for some effective mass anisotropy values. As shown in Fig. 8(a) and (c), for $m_{1}/m_{3}=1.6$ and 1.8, $b_{c2}(\theta,\phi,t)/b_{c2}(0,0,0)$ decreases monotonically with increasing $\theta$, but increases in Fig. 8(b), and for $\overline{m}_1/\overline{m}_3=0.6$, 0,8, and 1.0, the isotropic FS case, $b_{c2}(\theta,\phi,t)/b_{c2}(0,0,0)$ is a monotonically increasing function of $\theta$. For all of the cases plotted in Fig. 8, there is no peak in $b_{c2}(\theta,\phi,t)$ for $0 < \theta < \pi/2$, but it doesn't imply that $b_{c2}$ is a monotonic function of $\theta$ strictly. Recalling that there exits a crossing point in Fig. 3(b), we note that it doesn't stay monotonic near to the crossing point, but the change is slow and a little flat with increasing $\theta$.
\section{Conclusions}

In this paper we studied the temperature and angular dependence of the upper critical induction $B_{c2}(\theta,\phi,T)$ of $s$-wave and $d_{x^2-y^2}$-wave superconductors based
on the Gor'kov equation, which includes the anisotropic Zeeman energy arising from the ellipsoidal anisotropy of FS (with effective masses $m_i$ along the
three Cartesian coordinates for an orthorhomic crystal structure) that is treated self-consistently using the anisotropic Schr{\"o}diner-Pauli single particle Hamiltonian. While we
have derived the upper critical field for $s$-wave and $d$-wave superconductors with anisotropic effective masses, this work could  easily be applied to
any anisotropic pairing function. It is analogous to the treatment of $p$-wave superconductors with completely broken symmetry and ellipsoidal effective
mass symmetry, except for the addition of the Pauli pair-breaking effects in singlet-spin superconductors.

For an $s$-wave superconductor, we find the reduced upper critical induction $b_{c2}(\theta,\phi,t)$ is  angularly modulated by the universal orientation factor
$\alpha(\theta,\phi)$\cite{Klemm1980}, and the Pauli limiting can be exceeded by adjusting the effective mass, and the Zeeman energy almost can be neglected when comparing it with that for a $d_{x^2-y^2}$ OP. But it is not the case for $d_{x^{2}-y^{2}}$-wave superconductors, whose interaction and gap function depend upon the unit wave vector components in the $xy$ plane, which  become more complicated after the transformation of the single particle Hamiltonian to isotropic form.  $b_{c2}(\theta,\phi,t)$ for a $d_{x^2-y^2}$-wave superconductor not only changes along the axial $\phi$ direction with reduced temperature $t=T/T_c$, it shows an interesting $\phi$ dependence in the $xy$ plane. There is a $\pi/4$ shift in the $C_4$-symmetric calculated  $b_{c2}(\pi/2,\phi,t)$ patterns between low and high $t$ values when the Zeeman interaction is included. The effective mass anisotropy in the plane of the pairing interaction is significant, as for untwinned orthorhombic YBa$_2$Cu$_3$O$_{7-\delta}$, the $\phi$ dependence of $b_{c2}(\pi/2,\phi,t)$ in $xy$ plane at low $t$ changes from exhibiting $C_4$ symmetry  to $C_2$ symmetry. In addition, the variation of  $b_{c2}(\theta,0,t)$  with $\theta$ at low $t$ is a monotonic function of $\theta$.

\begin{acknowledgments}
This work was supported by the National Natural Science Foundation of China through Grant no. 11874083. A. Z. acknowledges financial support from the China Scholarship Council. R. A. K. was partially supported by the U. S. Air Force Office of Scientific Research (AFOSR) LRIR \#18RQCOR100, and the AFRL/SFFP Summer Faculty Fellowship Program provided by AFRL/RQ at WPAFB.
\end{acknowledgments}
\section*{Appendix}

In the following, we present  the forms of the recursion coefficients $C_{n,n'}$ for a $d_{x^2-y^2}$-wave superconductor.  In these expressions,
$\eta$ is given by Eq. (\ref{eta}) in the text,
and the $L_n^k(x)$ are the associated Laguerre polynomials.

\begin{eqnarray}
C_{n,n}&=&\pi T\sum_{\omega_{m}}\int_0^{\pi}\sin\theta_{k}d\theta_k\int_0^{\infty}d\xi e^{-2|\omega_{m}|\xi}e^{-\frac{1}{2}|\eta|^2}\nonumber\\
& &\times\cos(2\frac{e}{2 m_g}\alpha\tilde{B}\xi)
L_{n}^{0}(|\eta|^{2})\Bigl[\sin^{4}\theta_k\nonumber\\
& &\times(\frac{3A^2+3B^2}{8}+\frac{D^2+2AB}{8})+\sin^2\theta_k\cos^2\theta_k \nonumber\\
&&\frac{E^2+2AC+F^2+2BC}{2}+\cos^4\theta_k C^2\Bigr]- \frac{1}{N_{0}V_{0}},\nonumber\\
\end{eqnarray}
\begin{eqnarray}
C_{n,n+2}&=&\pi T\sum_{\omega_{m}}\int_0^{\pi}\sin\theta_{k}d\theta_k\int_0^{\infty}d\xi e^{-2|\omega_{m}|\xi}e^{-\frac{1}{2}|\eta|^2}\nonumber\\
& &\times\cos(2 \frac{e}{2 m_g}\alpha\tilde{B}\xi)
\frac{-|\eta|^2}{\sqrt{(n+2)(n+1)}}L_{n}^{2}(|\eta|^2)\Bigl[\sin^4\theta_k\nonumber\\
& &\times\Bigl(\frac{A^2-B^2}{4}-\frac{2AD+2BD}{8i}\Bigr)  \nonumber\\
&&+\frac{1}{4}\sin^{2}2\theta_k\nonumber\\
& &\times\Bigl(\frac{E^2+2AC-F^2-2BC}{4}-\frac{2EF+2CD}{4i}\Bigr)\Bigr],\nonumber\\
\end{eqnarray}
\begin{eqnarray}
C_{n,n-2}&=&\pi T\sum_{\omega_{m}}\int_0^{\pi}\sin\theta_{k}d\theta_k\int_0^{\infty}d\xi e^{-2|\omega_{m}|\xi}e^{-\frac{1}{2}|\eta|^2}\nonumber\\
& &\times\cos(2\frac{e}{2 m_g}\alpha\tilde{B}\xi)
\frac{-|\eta|^2}{\sqrt{n(n-1)}}L_{n-2}^{2}(|\eta|^2)\Bigl[\sin^4\theta_k\nonumber\\
& &\times\Bigl(\frac{A^2-B^2}{4}+ \frac{2AD+2BD}{8i}\Bigr) \nonumber\\
&&+\frac{1}{4}\sin^{2}2\theta_k\Bigl(\frac{E^2+2AC-F^2-2BC}{4}\nonumber\\
& &+\frac{2EF+2CD}{4i}\Bigr)\Bigr],
\end{eqnarray}
\begin{eqnarray}
C_{n,n+4}&=&\pi T\sum_{\omega_{m}}\int_0^{\pi}\sin\theta_{k}d\theta_k\int_0^{\infty}d\xi e^{-2|\omega_{m}|\xi}e^{-\frac{1}{2}|\eta|^2}\nonumber\\
& &\times\cos(2\frac{e}{2 m_g}\alpha\tilde{B}\xi)
\frac{(-|\eta|^2)^2}{\sqrt{(n+4)(n+3)(n+2)(n+1)}}\nonumber\\
& &\times L_{n}^{4}(|\eta|^2)\Bigl[\sin^4\theta_k\Bigl(\frac{A^2+B^2-D^2-2AB}{16}\nonumber\\
& &-\frac{AD-BD}{8i}\Bigr)\Bigr],
\end{eqnarray}
\begin{eqnarray}
C_{n,n-4}&=&\pi T\sum_{\omega_{m}}\int_0^{\pi}\sin\theta_{k}d\theta_k\int_0^{\infty}d\xi e^{-2|\omega_{m}|\xi}e^{-\frac{1}{2}|\eta|^2}\nonumber\\
& &\times\cos(2\frac{e}{2 m_g}\alpha\tilde{B}\xi)
\frac{(-|\eta|^2)^2}{\sqrt{n(n-1)(n-2)(n-3)}}\nonumber\\
& &\times L_{n-4}^{4}(|\eta|^2)\Bigl[\sin^4\theta_k   \nonumber\\
&&\times\Bigl(\frac{A^2+B^2-D^2-2AB}{16}+\frac{AD-BD}{8i}\Bigr)\Bigr],\nonumber\\
\end{eqnarray}
where
\begin{eqnarray}
A^2&=&\beta^4(\overline{m}_{1}\cos^2\phi'-\overline{m}_{2}\sin^2\phi')^2\cos^4\theta',\\
B^2&=&\beta^4(\overline{m}_{1}\sin^2\phi'-\overline{m}_{2}\cos^2\phi')^2\\
C^2&=&\beta^4(\overline{m}_{1}\cos^2\phi'-\overline{m}_{2}\sin^2\phi')^2\sin^4\theta',\\
D^2+2AB&=&\beta^4(\overline{m_{1}}+\overline{m}_{2})^2\cos^2\theta'\sin^{2}2\phi'\nonumber\\
& &+2\beta^4\cos^2\theta'(\overline{m}_{1}\cos^2\phi'-\overline{m}_{2}\sin^2\phi')\nonumber\\
& &\times(\overline{m}_{1}\sin^2\phi'-\overline{m}_{2}\cos^2\phi'),\\
E^{2}+2AC&=&\frac{3}{2}\beta^4\sin^{2}2\theta'(\overline{m}_{1}\cos^2\phi'\nonumber\\
& &-\overline{m}_{2}\sin^{2}\phi')^2,
\end{eqnarray}
\begin{eqnarray}
F^2+2BC&=&\beta^4\sin^2\theta'\sin^{2}2\phi'(\overline{m}_1+\overline{m}_2)^2\nonumber\\
& &+2\beta^4\sin^2\theta'\nonumber\\
& &\times\Bigl(\overline{m}_{1}\sin^2\phi'-\overline{m}_{2}\cos^2\phi'\Bigr)\nonumber\\
& &\times\Bigl(\overline{m}_{1}\cos^2\phi'-\overline{m}_{2}\sin^2\phi'\Bigr),\\
2AD&=&-2\beta^4\cos^3\theta'\sin 2\phi'(\overline{m}_{1}+\overline{m}_{2})\nonumber\\
& &\times\Bigl(\overline{m}_{1}\cos^2\phi'-\overline{m}_{2}\sin^2\phi'\Bigr),
\end{eqnarray}
\begin{eqnarray}
2BD&=&-2\beta^4\cos\theta'\sin 2\phi'(\overline{m}_{1}+\overline{m}_{2})\nonumber\\
& &\times\Bigl(\overline{m}_{1}\sin^2\phi'-\overline{m}_{2}\cos^2\phi'\Bigr),
\end{eqnarray}
\begin{eqnarray}
2EF+2CD&=&-3\beta^4\sin\theta'\sin 2\theta'\sin 2\phi'\nonumber\\
& &\times(\overline{m}_{1}+\overline{m}_{2})\nonumber\\
& &\times\Bigl(\overline{m}_{1}\cos^2\phi'-\overline{m}_{2}\sin^2\phi'\Bigr).
\end{eqnarray}

\end{document}